\documentclass[sigplan,screen]{acmart}
\AtBeginDocument{%
  \providecommand\BibTeX{{%
    \normalfont B\kern-0.5em{\scshape i\kern-0.25em b}\kern-0.8em\TeX}}}

\setcopyright{acmcopyright}
\copyrightyear{2026}
\acmYear{2026}
\acmDOI{XXXXXXX.XXXXXXX}

\acmConference[Conference acronym 'XX]{Make sure to enter the correct
  conference title from your rights confirmation emai}{June 03--05,
  2026}{ }

\acmPrice{15.00}
\acmISBN{978-1-4503-XXXX-X/18/06}

\usepackage{amsmath,amsfonts}
\usepackage{algorithmic}
\usepackage{graphicx, subfigure, verbatim, url,pifont}
\usepackage{stmaryrd}
\usepackage[ruled,vlined]{algorithm2e}

 \usepackage{multirow}
 \usepackage{booktabs}
\usepackage{makecell}

\begin{document}

\title{R2E-VID: Two-Stage Robust Routing via Temporal Gating for Elastic Edge-Cloud Video Inference}



\author{Zheming Yang\textsuperscript{\rm 1},
    Lulu Zuo\textsuperscript{\rm 1,2,3},
    Shun Lu\textsuperscript{\rm 5},
    Yangyu Zhang\textsuperscript{\rm 1,2},
    Zhicheng Li\textsuperscript{\rm 1,2},
    Xiangyang Li\textsuperscript{\rm 1}, and
    Yang You\textsuperscript{\rm 4}\\}
\affiliation{%
  \textsuperscript{\rm 1}\institution{Institute of Computing Technology, Chinese Academy of Sciences}
\textsuperscript{\rm 2}\institution{University of Chinese Academy of Sciences}
\textsuperscript{\rm 3}\institution{Peng Cheng Laboratory}
 \textsuperscript{\rm 4}\institution{National University of Singapore}
 \textsuperscript{\rm 5}\institution{Beijing Kuaishou Technology Co., Ltd.}
 \textcolor{white}{ \country{China}}
  }









\renewcommand{\shortauthors}{Anonymous Authors.}

\begin{abstract}
With the rapid growth of large-scale video analytics applications, edge–cloud collaborative systems have become the dominant paradigm for real-time inference. However, existing approaches often fail to dynamically adapt to heterogeneous video content and fluctuating resource conditions, resulting in suboptimal routing efficiency and excessive computational costs. In this paper, we propose R2E-VID, a two-stage robust routing framework via temporal gating for elastic edge–cloud video inference. In the first stage, R2E-VID introduces a temporal gating mechanism that models the temporal consistency and motion dynamics of incoming video streams to predict the optimal routing pattern for each segment. This enables adaptive partitioning of inference workloads between edge and cloud nodes, achieving fine-grained spatiotemporal elasticity. In the second stage, a robust routing optimization module refines the allocation through multi-model adaptation, jointly minimizing inference delay and resource consumption under dynamic network and workload variations. Extensive experiments on public datasets demonstrate that R2E-VID achieves up to 60\% reduction in overall cost compared to cloud-centric baselines, and delivers 35–45\% lower delay while improving inference accuracy by 2–7\% over state-of-the-art edge–cloud solutions.
\end{abstract}






\maketitle

\section{Introduction}
\label{sec:intro}

With the rapid development of technologies such as deep learning and the Internet of Things (IoT), the deployment and application of various IoT devices are becoming increasingly widespread \cite{ji2020crowd515151}, resulting in the inference tasks generated increasing dramatically. Many cities around the world have deployed millions of cameras at traffic intersections, communities, and other places \cite{hsieh2018focus121212}, which are used for inference tasks (e.g., object detection task for a picture or a video stream) of various new scenes \cite{talker2024mind101, 102102102}. However, many inference tasks have extremely high requirements for accuracy and delay in practical applications \cite{li2019edge636363}. A large number of deep neural network (DNN) models are deployed on servers for performing inference tasks \cite{shao2020communication646464, cetinkaya2024ranked103}. It is difficult to satisfy low delay requirements if all inference tasks are uploaded on the cloud due to limited bandwidth resources \cite{chen2019deep151515,100,200}. To compensate for the above shortcomings, edge computing \cite{shi2016edge161616,300} has a great advantage in saving transmission bandwidth due to being more distributed and closer to the data source. However, the limited computation resources of edge servers can usually only handle some simple inference tasks \cite{zhou2019edge171717,yang2023visual656565}, and it is difficult to meet the accuracy requirements of complex tasks.

\begin{figure}[t!]
	\centering 
		\includegraphics[width=1\linewidth]{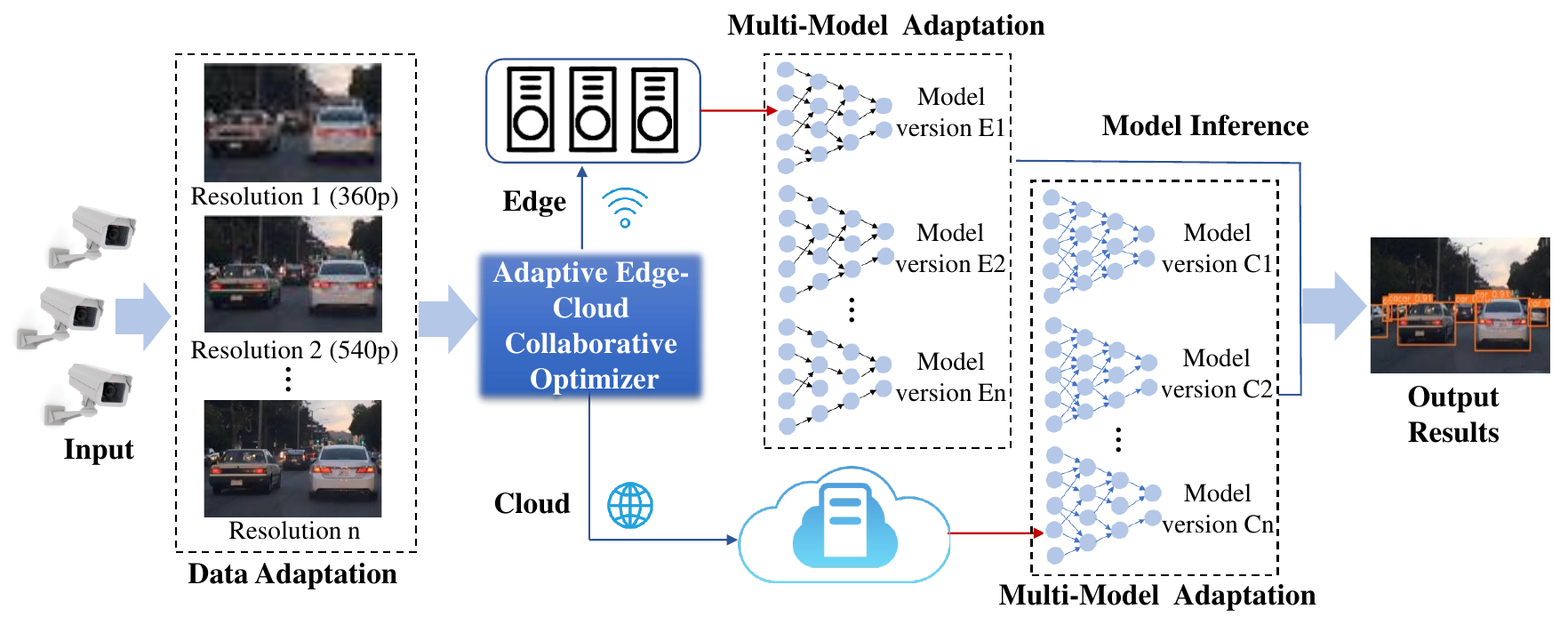}
	\caption{The illustration of edge-cloud collaborative architecture for video inference.}
	\label{figure1}
\end{figure}




\par
Recent studies have proposed various task allocation mechanisms for edge-cloud collaboration systems \cite{su2022prediction363636,434343,444444,454545,li2023moby565656,wang2023shoggoth575757}, as shown in Figure~\ref{figure1}. Efficient processing of heterogeneous tasks necessitates not only dynamic model selection but also optimal data selection  \cite{koupil2022mm535353,1100}. While large models are characterized by high inference accuracy, they are also associated with significant delay and energy consumption\cite{600,700}. Conversely, smaller models offer reduced inference delay and energy consumption but at the cost of lower accuracy \cite{800,900}. To mitigate costs, deploying models of varying sizes across servers can cater to diverse inference tasks effectively \cite{yuan2022mlink525252}. In practical scenarios, task requirements frequently fluctuate with changes in time and environment \cite{kim2023moca595959,1200}. For instance, some simple tasks might be assigned unnecessarily complex models or erroneously offloaded to cloud servers, incurring avoidable costs \cite{434343,400}. This complexity presents a substantial challenge for edge-cloud collaboration and adaptive inference using multiple models under resource-constrained conditions \cite{xu2018scaling494949,500}.

\par
In this paper, we propose a cost-efficient elastic inference framework with two-stage robust optimization for edge-cloud video analysis, named R2E-VID. It can adaptively adjust the resolution, frame rate, and model version according to different task requirements, and decide whether to transfer it to the cloud or the edge. The goal is to reduce costs while meeting the accuracy requirements. This work emphasizes multi-dimensional collaborative and complex optimization in efficient inference, providing a joint optimization scheme for the tradeoff between accuracy and cost. The main contributions of this paper can be summarized as follows.

\par

\begin{itemize}
\item We develop R2E-VID, a two-stage robust optimization framework for elastic edge–cloud video inference. It decouples the decision process into adaptive video configuration and robust model selection, enabling fine-grained accuracy and cost tradeoffs under dynamic network and resource conditions. 

\item We propose a temporal gating–based routing module that captures temporal consistency and motion dynamics in video streams. It enables the system to dynamically partition inference workloads across edge and cloud nodes in response to content variations. 

\item We evaluate the performance of the framework and compare it with mainstream baseline methods. Experimental results on the public dataset demonstrate that our method has the best success rates for meeting the task requirements. It can reduce the inference cost by 35\%-60\% and without sacrificing accuracy. 
\end{itemize}

\section{Motivation Analysis}
\begin{figure}[t!]
	\centering 
    \subfigure[540p]{
    \label{Fig.sub.1.1}
    \includegraphics[width=0.1\textwidth]{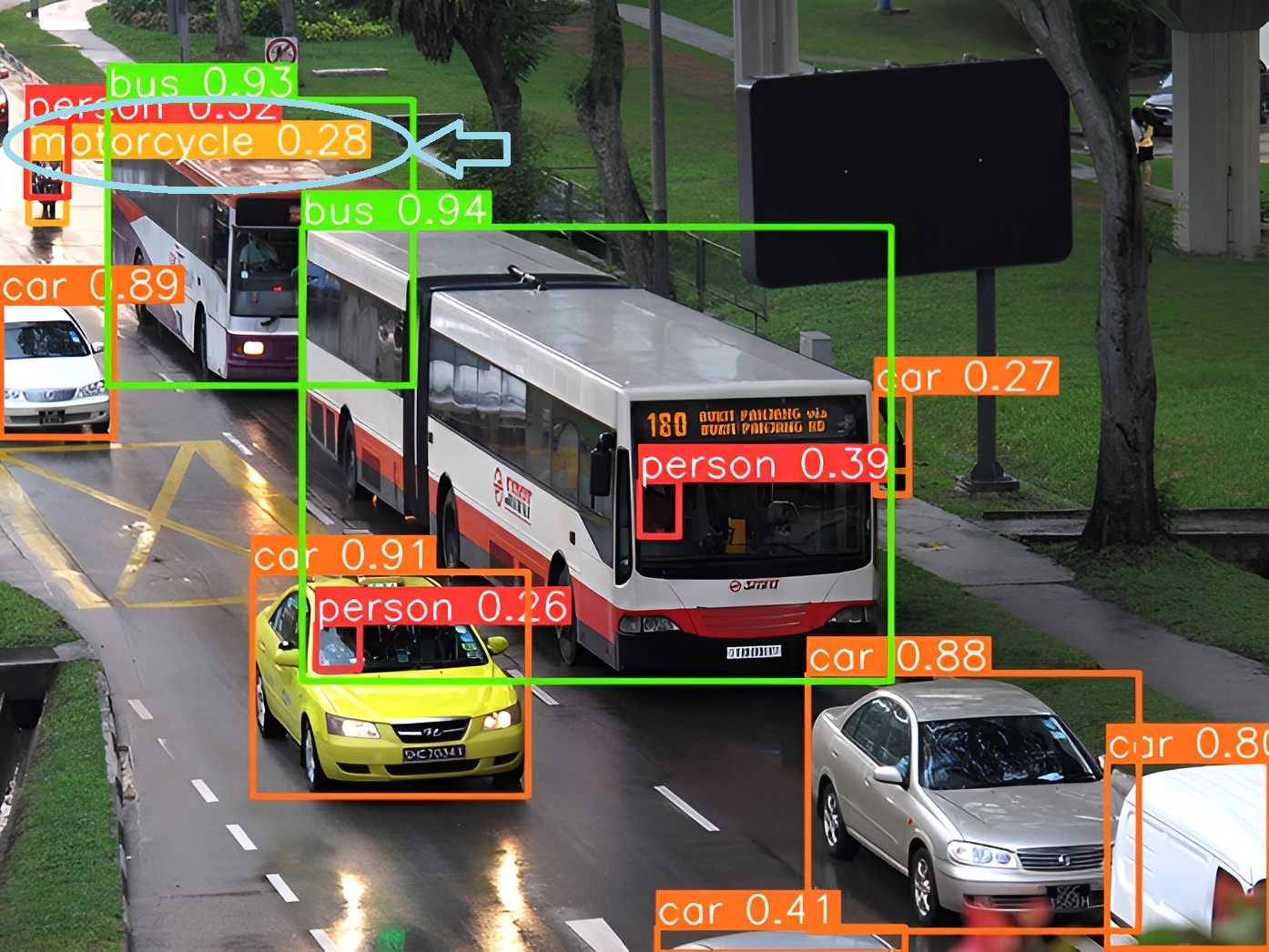}}
    \subfigure[720p]{
    \label{Fig.sub.1.2}
    \includegraphics[width=0.1\textwidth]{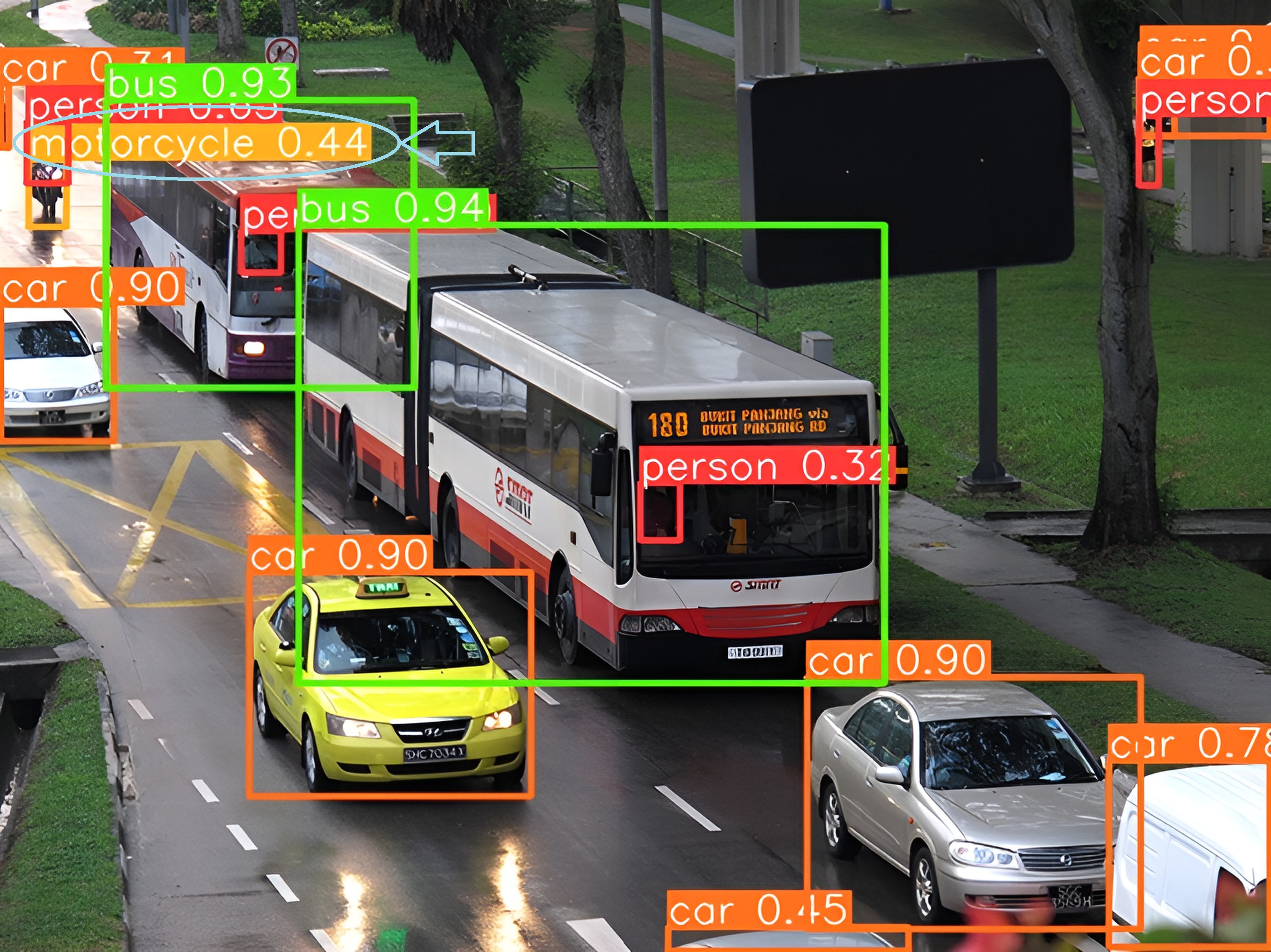}}
    \subfigure[900p]{
    \label{Fig.sub.1.3}
    \includegraphics[width=0.1\textwidth]{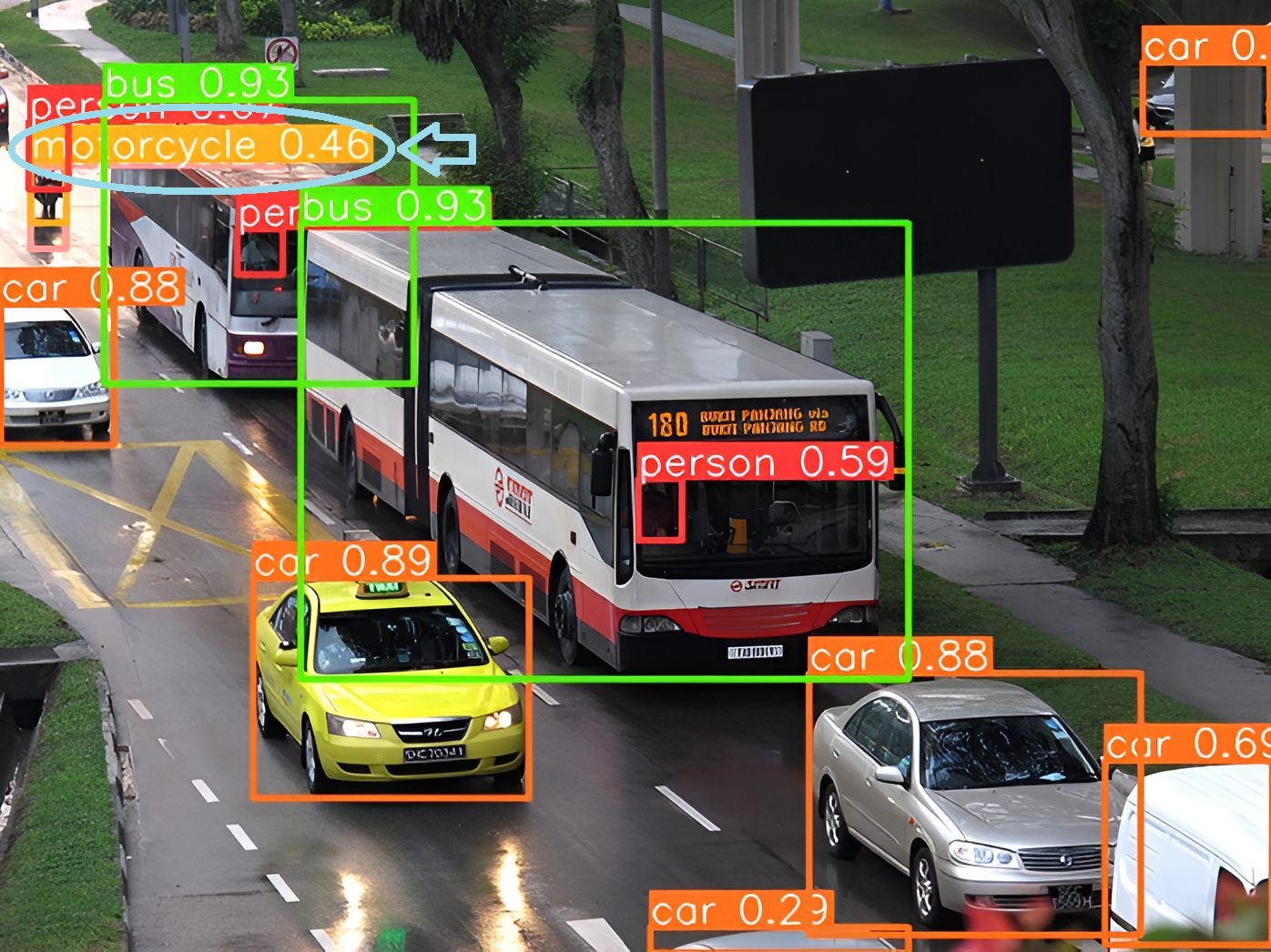}}
    \subfigure[1080p]{
    \label{Fig.sub.1.4}
    \includegraphics[width=0.1\textwidth]{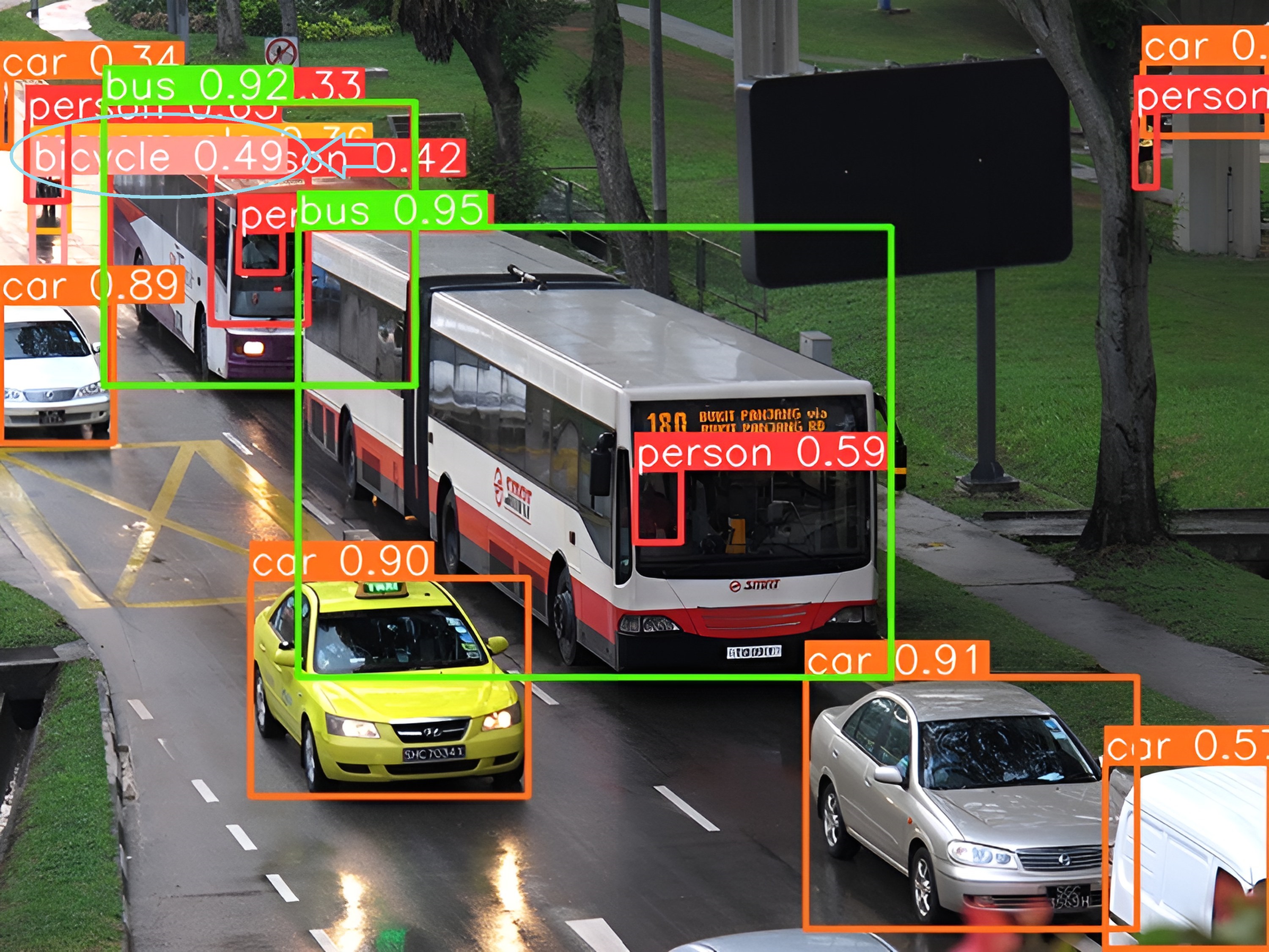}}
    
    \subfigure[Edge, small models]{
    \label{Fig.sub.8.1}
    \includegraphics[width=0.2\textwidth]{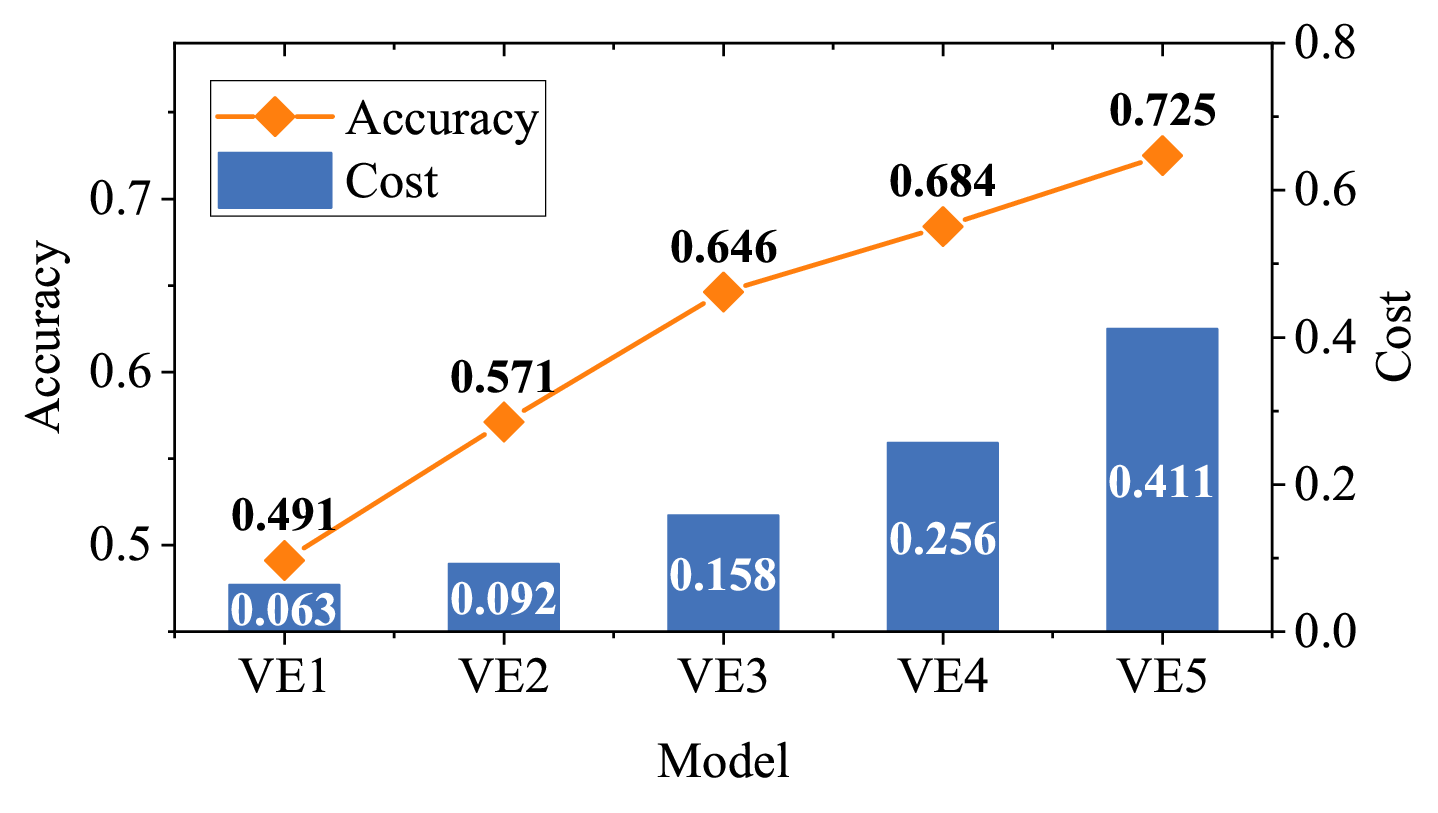}}  
    \subfigure[Cloud, complex models]{
    \label{Fig.sub.8.2}
    \includegraphics[width=0.2\textwidth]{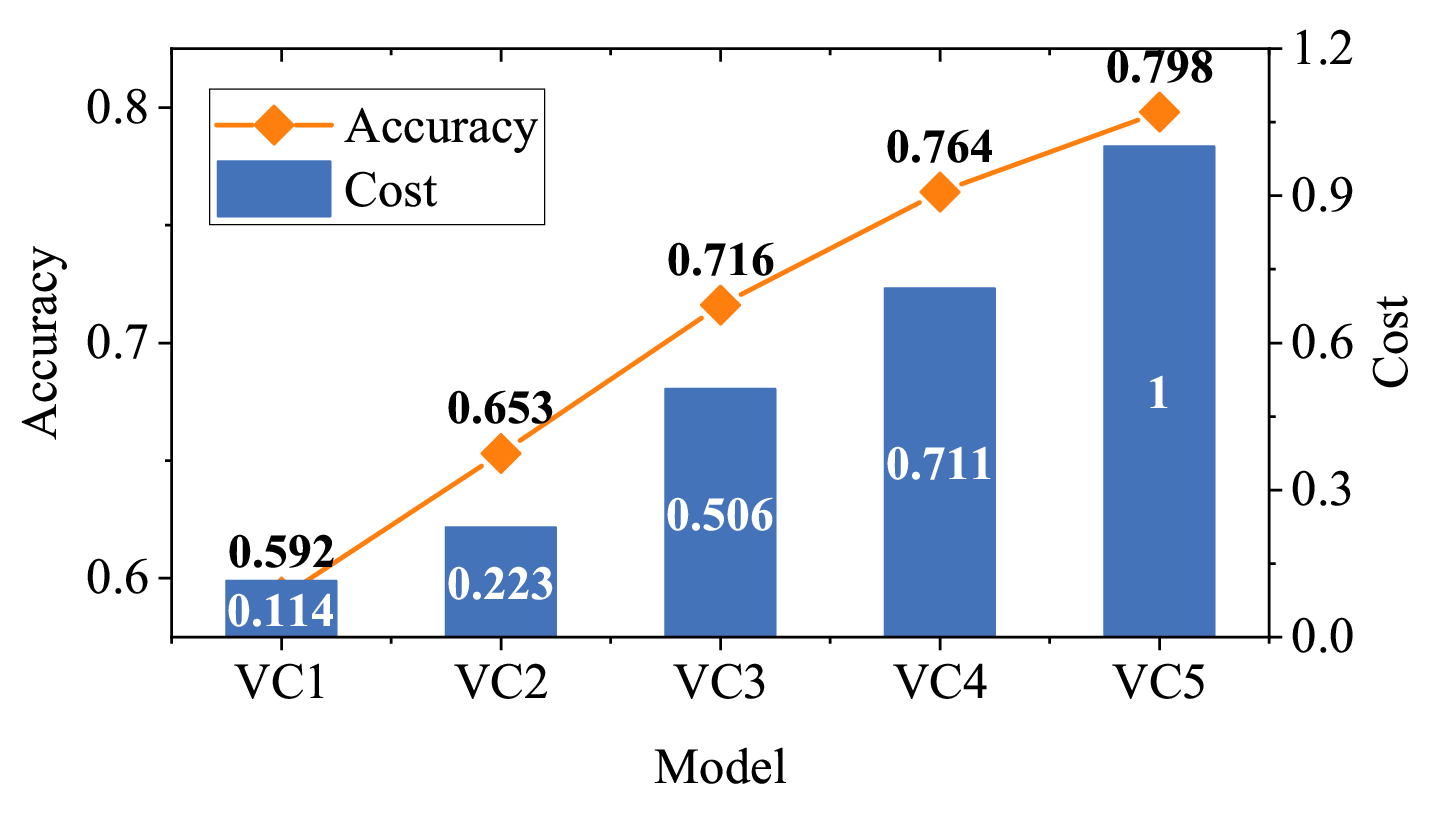}}
    \caption{The (a), (b), (c), and (d) are the inference results for different resolutions. The (e) and (f) are the results for different models on edge server and cloud server. The normalized delay and energy consumption are multiplied and then summed to obtain the cost.}
	\label{figure2}
\end{figure}

The execution of real-time video analytics in edge–cloud environments fundamentally hinges on jointly optimizing data fidelity and model complexity \cite{zhang2017live404040, 484848, 1000}. To illustrate this challenge, we evaluate multiple model variants with heterogeneous computational footprints and test four input resolutions across both edge and cloud servers. As shown in Figure~\ref{figure2}, these results reveal a non-linear and highly coupled tradeoff between accuracy, delay, and computational cost. More critically, the optimal configuration varies drastically across video segments due to temporal changes in motion dynamics, scene complexity, and resource fluctuations \cite{jiang2018chameleon444}. As prior studies note, deploying diverse model variants across edge and cloud tiers can effectively balance efficiency and quality \cite{koupil2022mm535353, yuan2022mlink525252}, but doing so requires precise coordination between routing, model selection, and resolution adaptation. Without such coordination, misconfigurations accumulate and degrade system performance. These observations motivate the need for a unified, temporally adaptive optimization framework capable of jointly determining video configuration (e.g., resolution, frame rate), edge–cloud partitioning, and model selection. This is precisely the gap that R2E-VID aims to address through its two-stage robust routing design.


\section{The Proposed R2E-VID Framework}
\label{method}

In this section, we introduce the R2E-VID framework, which is designed for a two-stage robust optimization paradigm. As illustrated in Figure~\ref{figure3}, the framework is composed of two tightly coupled optimization stages that jointly address the challenges of dynamic task requirements and heterogeneous system resources. In the first stage, R2E-VID leverages adaptive edge-cloud collaborative configuration to determine the optimal task resolution, frame rate, and server assignment (edge or cloud). In the second stage, the framework performs multi-model elastic inference, dynamically selecting the most appropriate model version based on the initial configuration and real-time resource conditions. This ensures that the inference process remains both cost-efficient and accurate under varying workloads.

\subsection{Two-Stage Robust Optimization Formulation}

We consider a collection of video inference tasks generated by end devices, denoted as $\mathcal{T}={t_1,t_2,\dots,t_M}$. Each task must be routed to either an edge server or a cloud server, captured by the binary variable $y_i\in{0,1}$, where $y_i=0$ indicates edge processing and $y_i=1$ indicates cloud offloading. To support elastic video uploading, the system can select one resolution from $\mathcal{R}={r_1,\dots,r_N}$ and one frame rate from $\mathcal{P}={p_1,\dots,p_Z}$. Meanwhile, inference is performed using a model version chosen from $\mathcal{V}={v_1,\dots,v_K}$, where each version represents a distinct accuracy–cost trade-off. For each video task, the system jointly decides (i) the offloading destination, (ii) the resolution–frame rate pair for transmission, and (iii) the model version used for inference. These decisions must satisfy task-specific accuracy requirements while minimizing the end-to-end processing cost, which integrates both delay $D_i$ and energy consumption $E_i$ using the tradeoff parameter $\beta$. The optimization problem is formulated as follows:

\begin{figure}[t!]
	\centering 
		\includegraphics[width=1\linewidth]{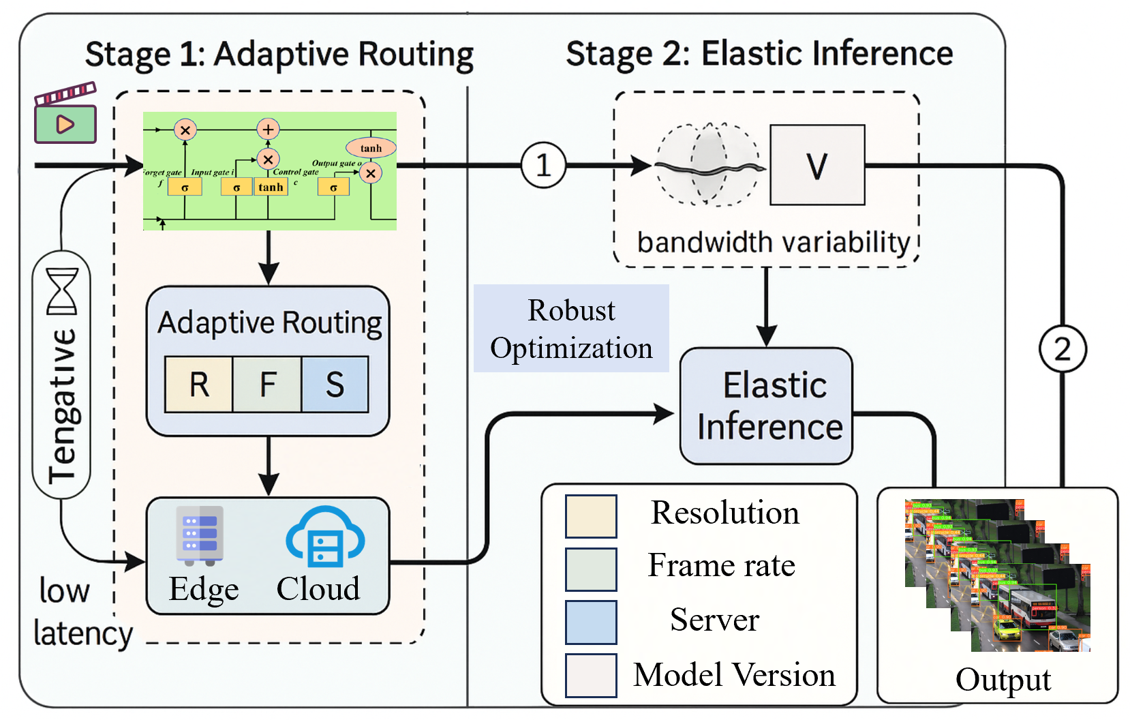}
	\caption{The workflow of the proposed R2E-VID framework.}
	\label{figure3}
\end{figure}

\begin{equation}
\begin{array}{ll}
 &  \min \quad  \sum_{i=1}^{M}\left(D_{i}+\beta E_{i}\right) \\

\text { s.t. } &  C_{1}: f_{i}(r,v,z) \geq A_{i}^{q},i\in \left\{1, 2, \cdots M\right\} \\

& C_{2}: x_{i}=\{0,1\},i\in \left\{1, 2, \cdots M\right\}\\

& C_{3}: \sum_{n=1}^{N} x_{i, n}=1, x_{i, n}=\{0,1\}, r_{n} \in \mathcal{R}  \\

& C_{4}: \sum_{z=1}^{Z} x_{i, z}=1, x_{i, z}=\{0,1\}, p_{z} \in \mathcal{P}  \\

& C_{5}: \sum_{k=1}^{K} x_{i, k}=1, x_{i, k}=\{0,1\},v_{k} \in \mathcal{V} \\

& C_{6}: \sum_{i=1}^{M} B_{i}\leq B

\end{array}
\end{equation}

The $f_{i}(r,v,z)$ is the accuracy function,  $A_{i}^{q}$ is the accuracy requirement. $x_{i}$ is a binary variable used to ensure that each task can only select one resolution, frame rate, and model version. $B$ is the network bandwidth. Because this problem couples discrete routing/model-selection decisions with continuous delay and energy cost terms. This makes direct optimization intractable in real-world systems, especially under long-term dynamics and fluctuating resource conditions.

Building on the above formulation, we cast edge–cloud video configuration and multi-model selection as a two-stage robust optimization problem with sequential decisions under uncertainty. In the first stage, the system adaptively chooses input resolution, frame rate, and execution location according to accuracy needs and network conditions, defining the feasible space for the second stage. After observing actual resources, network variations, and intermediate results, the second stage selects an appropriate model version to minimize end-to-end cost. Because offloading, model choice, and uncertain execution are tightly coupled, the full problem is intractable. Thus, we adopt a decomposition-based robust optimization approach that converts the high-dimensional MINLP into two tractable subproblems \cite{zeng2013solving282828, goerigk2022two262626}. Formally, the model is expressed as:

\begin{equation}
\begin{array}{ll}
 &  \min \limits_{y} \mathbf{c}^{T} y+\max \limits_{u \in \mathcal{U}} \min \limits_{v \in F(y, u)} \mathbf{b}^{T} v \\

\text { s.t. } & \qquad C_{1}, C_{2}, C_{3}, C_{4}, C_{5}, C_{6}

\end{array}
\end{equation}
where $y$ is the first-stage decision variable and $v$ is the second-stage decision variable. The uncertainty set $\mathcal{U}$ captures environmental and task-related uncertainties. The $F(y, u)= \left\{\mathbf{v} \in \mathbf{\mathcal{S}}_{\mathbf{v}}: \mathbf{G} \mathbf{v} \geq \mathbf{h}-\mathbf{Q} y-\mathbf{L} u\right\}$ denotes the second stage problem, and $\mathcal{S}$ is a polyhedron. We use $\mathbf{c}^{T}$ to denote the cost matrix with different resolutions, frame rate, and servers, and $\mathbf{b}^{T}$ is the cost matrix with different models. This two-stage formulation explicitly accounts for system uncertainty and decomposes the complex joint optimization into a structured decision pipeline.


\subsection{Stage 1: Adaptive Edge-Cloud Configuration}

For Eq. (2), we decompose the original problem into two subproblems with mixed integer linear characteristics through the Benders decomposition algorithm. Since the model selection problem of the second stage is a linear programming (LP) problem for $v$, it is feasible for any given $y$ and $u$. Then let $\pi$ be its dual variable and merge it with the maximization over $u$. Finally, the subproblem 1 in the Benders-dual \cite{rahmaniani2020benders616161} method is obtained:

\begin{equation}
\begin{aligned}
 & \mathcal{SP}_{1}:  \mathcal{Q}(y)= & \max _{u, \pi}\left\{(\mathbf{h}-\mathbf{Q}y-\mathbf{L} u)^T \pi: \mathbf{G}^T \pi \leq \mathbf{b}\right. \}
\end{aligned}
\end{equation}
where $\mathbf{Q}$ is the first stage coefficient matrix, $\mathbf{L}$ is the second-stage coefficient matrix. $\mathbf{G}$ is the total stage coefficient matrix, and $\mathbf{h}$ is the accuracy demand vector. The objective function is converted from the original minimization to the maximization of the dual variable $\pi$. And then combined with $u$ to form a bilinear optimization problem. Assume that the optimal solution of subproblem 1 is $\left(u_{\mathrm{i}}^*, \pi_{\mathrm{i}}^*\right)$ for a given $y_i^*$. According to the duality theorem, the following cutting planes can be constructed: $\eta \geq\left(\mathbf{h}-\mathbf{Q} y-\mathbf{L} \mathbf{u}_i^*\right)^{\mathrm{T}} \pi_i^*$.  The $\eta=\max _{u \in \mathcal{U}} \min _{v \in F(y, u)} \mathbf{b}^T v$ is a one-dimensional scalar, and then the cut-plane constraint is added to the first stage of optimization to obtain the master problem 1: 

\begin{equation}
\begin{array}{ll}
 & \mathcal{MP}_{1}:   \min \limits_{y} \quad \mathbf{c}^{T} y + \eta \\

\text { s.t. } & \quad C_{1}, C_{2}, C_{3}, C_{4}\\
& \eta \geq\left(\mathbf{h}-\mathbf{Q}y-\mathbf{L} \mathbf{u}_i^*\right)^{\mathrm{T}} \pi_i^*, y \in \mathcal{S}_{y}

\end{array}
\end{equation}

\begin{figure}[t!]
	\centering 
		\includegraphics[width=1\linewidth]{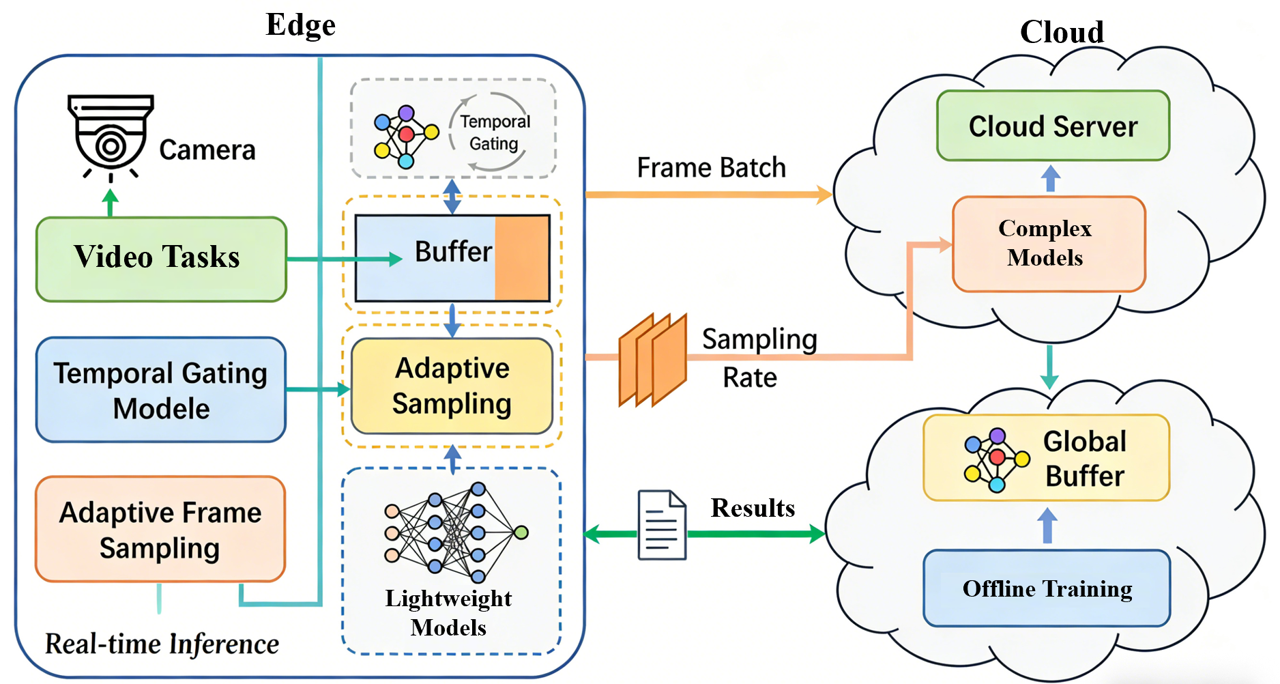}
	\caption{The illustration of adaptive edge-cloud collaborative configuration via temporal gating.}
	\label{figure4}
\end{figure}

In the first stage, the high-level master problem \(\mathcal{MP}_1\) determines the binary first-stage decision variable \(y\) and an auxiliary variable \(\eta\) that serves as a piecewise-linear underestimator of the second-stage value function. The variable \(y\) encodes task-level configuration policies across edge and cloud nodes, including input resolution, sampling rate, and the binary delegation of each video segment. To enable content-aware initialization and constraint generation, we introduce a temporal gating mechanism that captures both short-term motion dynamics and long-range temporal consistency, as shown in Figure~\ref{figure4}. Unlike conventional approaches that treat each frame independently or rely on fixed sampling patterns, our gating module adaptively determines where and how often to trigger cloud offloading based on video content volatility, as shown in Algorithm 1.

Given an incoming video segment \(\mathcal{S}_t = \{I_{t-K+1}, \dots, I_t\}\) consisting of \(K\) consecutive frames, we first compute a frame-wise difference representation \(\Delta x_t = \phi(I_t, I_{t-1})\), where \(\phi\) is a lightweight operation combining pixel-wise absolute difference and histogram-based motion magnitude. To suppress noise, we apply a spatial downsampling factor of \(4\times\) and a temporal moving average over a small window of size \(3\). The resulting \(\Delta x_t \in \mathbb{R}^d\) captures local motion intensity. The temporal gating unit is implemented as a gated recurrent cell with content-adaptive forget bias:

\begin{equation}
g_t = \sigma\left(W_g \Delta x_t + U_g h_{t-1} + b_g + \alpha \cdot \text{Var}(\Delta x_{t-T:t})\right)
\end{equation}

Here, \(h_{t-1} \in \mathbb{R}^m\) is the hidden state summarizing historical motion patterns, and the additional term \(\alpha \cdot \text{Var}(\Delta x_{t-T:t})\) introduces a \textit{volatility modulation}: when the recent \(T\)-frame motion variance exceeds a threshold, the gate opens more aggressively to prevent missed critical events. The parameters \(W_g, U_g, b_g, \alpha\) are learned via a meta-training procedure described below. The hidden state evolves as:

\begin{equation}
h_t = (1 - g_t) \odot h_{t-1} + g_t \odot \tanh\left(W_h \Delta x_t + U_h (r_t \odot h_{t-1}) + b_h\right)
\end{equation}
with a reset gate \(r_t = \sigma(W_r \Delta x_t + U_r h_{t-1} + b_r)\). The output of this module is a \textbf{temporal significance score} \(\tau_t = \sigma(W_o h_t + b_o) \in [0,1]\), which quantifies the expected benefit of cloud assistance for segment \(\mathcal{S}_t\).

\begin{algorithm}[t]
\textbf{Input}: Task set $\mathcal{T}$, Resolution set $\mathcal{R}$, Threshold $A_{i}^{q}$, Video segment $\mathcal{S}_t$ \\
\textbf{Output}: First-stage configuration $r_{n}^*$, Decision $y_{i}^*$, Temporal gate $g_t$ \\
\textbf{Procedure}: \\
1: \textbf{for} $i=1$ to $M$, $n=1$ to $N$ \textbf{do} \\
2: \qquad Compute motion feature $\Delta x_t$ and temporal gate $g_t$ (content-adaptive) \\
3: \qquad Calculate $f_{i}(r,v_1)$ guided by temporal significance score $\tau_t$ \\
4: \qquad \textbf{if} $f_{i}(r,v_1) \geq A_{i}^{q}$ \textbf{then} $r_{n}^* \leftarrow r_n$ \\
5: \qquad \textbf{else} $r_n \leftarrow$ higher resolution, update gate $g_t$ \\
6: \qquad Enforce temporal consistency constraint $\|y_t - y_{t-1}\|_1 \le \delta$ \\
7: \textbf{end for} \\
8: \textbf{while} $f_{i}(r,v_1) < A_{i}^{q}$ \textbf{do} $y_i^* \leftarrow 1$ (cloud offloading) \\
9: \textbf{end while} \\
10: \textbf{return} $r_{n}^*, y_i^*, g_t$ \\
\caption{Adaptive Edge-cloud Collaborative Configuration Algorithm via Temporal Gating}
\end{algorithm}

The gating module is trained via a two-stage curriculum: offline warm-up on diverse video categories minimizes \(\mathcal{L}_{\text{acc}} + \lambda_1 \mathcal{L}_{\text{lat}} + \lambda_2 \mathcal{L}_{\text{comp}}\), followed by online fine-tuning with a proximal regularizer to prevent catastrophic forgetting. The output configurations serve as a warm-start for \(\mathcal{MP}_1\), and we further impose a temporal consistency constraint \(\|y_t - y_{t-1}\|_1 \le \delta(|\tau_t - \tau_{t-1}|)\) to prevent oscillatory switching between edge and cloud. The resulting first-stage plan provides a robust, temporally coherent configuration that aligns content dynamics with heterogeneous infrastructure, forming an optimized feasible region for second-stage model selection.


\subsection{Stage 2: Multi-Model Elastic Inference}
However, the result presented in \textbf{Stage 1} represents merely an initial phase of the comprehensive analysis required to address the complexity of the problem at hand. If only the decision variables and constraints in the first stage are considered, the current optimization result can be regarded as a relaxed version of the whole problem. We need to be further optimized in combination with the subproblem of the second stage. Similar to  Eq. (3), according to the Benders-dual method, the expression of subproblem 2 is as follows: 

\begin{equation}
\begin{aligned}
 & \mathcal{SP}_{2}:  \mathcal{Q}(y)=\left\{\max _{u \in \mathcal{U}} \min _{\mathbf{v}} \mathbf{b}^T v: \mathbf{G}v \geq \mathbf{h}-\mathbf{Q}y-\mathbf{L} u \right\}
\end{aligned}
\end{equation}

Similar to the Benders-dual method, we can either derive an optimal solution with a finite optimum $\mathbf{Q} y$ or identify some $u \in \mathcal{U} $ for which the second stage decision problem is infeasible. Then, the $\mathbf{Q} y$ under the infeasible case is set to $+\infty$. The expression of master problem 2 is as follows: 

\begin{equation}
\begin{array}{ll}
 &   \mathcal{MP}_{2}:  \max \limits_{u \in \mathcal{U}} \min \limits_{v \in F(y, u)} \mathbf{b}^{T} v \\
\text { s.t. } & \qquad \quad C_{4}, C_{5}.\\
& \mathbf{G}v \geq \mathbf{h}-\mathbf{Q}y-\mathbf{L} u, v \in \mathcal{S}_{v}

\end{array}
\end{equation}

\par
According to the form of Eq. (2), it can be found that the objective functions and constraints of Eq. (4) and Eq. (8) can be coupled with each other. It can be proved that the original problem is equivalent to the two master problems \cite{takriti2004robust474747}. Since master problem 2 is a max-min problem, it is difficult to solve it directly using existing algorithms. Therefore, we first introduce the uncertainty set $\mathcal{U}$ of the objective function, expressed as follows:

\begin{equation}
\mathcal{U}=\left\{u: u_{k}=\underline{u}_{k}+g_{k} \tilde{u}_{k}, g_{k} \in[0,1], \sum_{k} g_{k} \leq \Gamma\right\}
\end{equation}
where $\underline{u}_{k}$ is the basic target cost demand, $\tilde{u}_{k}$ is the maximum deviation of the target cost, and $g_{k}$ is a decision variable. $\Gamma$ is a predefined integer value introduced to control the uncertainty constraint of the target cost. Based on the strong dual theory \cite{li2006towards272727}, we perform a dual transformation on the model of master problem 2. By turning the inner min problem into a max problem and merging it with the outer max problem, we obtain the following dual problem:

\begin{equation}
\begin{array}{ll}
 &   \max \limits_{u, \lambda, \pi} \quad \sum_{k}\left[u_{k}+g_{k}\tilde{u}_{k} \right] \lambda_{i}-\sum_{i} \pi_{i} \varphi_{ik} \\
\text { s.t. } &  \lambda_{i}-\pi_{i} \leq \varphi_{ik}\\
&  \lambda_{i} \in R^{+}, \pi_{i} \in R^{+}
\end{array}
\end{equation}

Let $\pi$ and $\lambda$ denote the dual variables generated by the two introduced constraints. According to the work in \cite{bertsimas2012adaptive303030}, the optimal solution to this dual problem corresponds to $u$ as a pole of the uncertainty set $\mathcal{U}$. When Eq. (10) takes the maximum value, the value of the uncertain variable $u$ should be the boundary of the interval of Eq. (9).

\begin{algorithm}[t]
\caption{Multi-Model Elastic Inference Acceleration Algorithm based on Robust Optimization}
\KwIn{Model set $\mathcal{V}$; initial config $(r_n^*, p_z^*, y_i^*)$; uncertainty set $\mathcal{U}$}
\KwOut{Final configuration $(r_n, p_z, y_i, v_k)$}

Initialize $O_{up}\!\leftarrow\!+\infty$, $O_{down}\!\leftarrow\!-\infty$; select initial $u_0\!\in\!\mathcal{U}$\;

\While{iteration $< T$}{
    \For{$k = 1$ \KwTo $K$}{
        Compute first-stage solution $y_i$ via $\mathcal{MP}_1$ under $u_w$; update
        $O_{down} \leftarrow \mathbf{c}^T y_i$\;

        Solve second-stage $\mathcal{MP}_2$ to obtain $v_k$; update
        $O_{up} \leftarrow \min(O_{up},\, \mathbf{c}^T y_i + \mathbf{b}^T v_k)$\;

        \If{$O_{up} - O_{down} \le \theta$}{
            $\mathcal{F}^* \leftarrow O_{up}$; break\;
        }
        Update uncertainty sample: $u_{w}\leftarrow u_{w+1}$\;
    }
}
\Return{$(r_n, p_z, y_i, v_k)$}\;
\end{algorithm}

\par
Then, we propose a multi-model elastic inference acceleration algorithm based on robust optimization to solve master problem 2, as shown in Algorithm 2. It first dynamically generates the constraints of decision variables in the original space \cite{zeng2013solving282828}, and then iteratively optimizes the target value according to the uncertainty set $\mathcal{U}$ and the results in the first stage. In each iteration, for each constraint of the $\mathcal{MP}_{1}$, fix the master variables and solve the corresponding $\mathcal{MP}_{2}$. The optimal auxiliary variable columns are obtained and added to the set of columns. Then, update the objective function and constraints for $\mathcal{MP}_{2}$, including the newly added auxiliary variable columns, and increase the number of iterations. This process is repeated until the termination criterion is satisfied. Finally, the current result is enhanced by progressively generating columns of auxiliary variables, leading to an approximate solution for the edge-cloud collaborative configuration and model selection.


\section{Performance Evaluation}  \label{Performance Evaluation}

\begin{figure*}[th]
	\centering 
    \subfigure[COCO]{
    \label{Fig.sub.8.1}
    \includegraphics[width=0.32\textwidth]{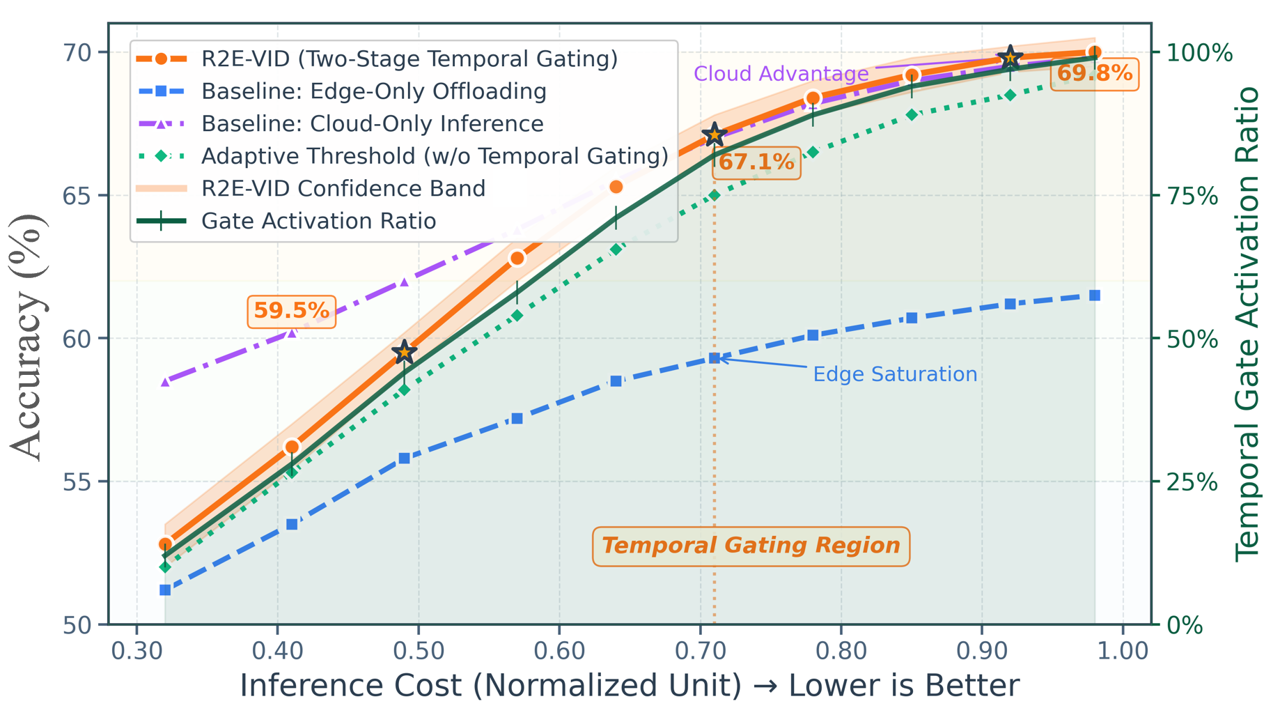}}
    \subfigure[UA-DETRAC]{
    \label{Fig.sub.8.2}
    \includegraphics[width=0.32\textwidth]{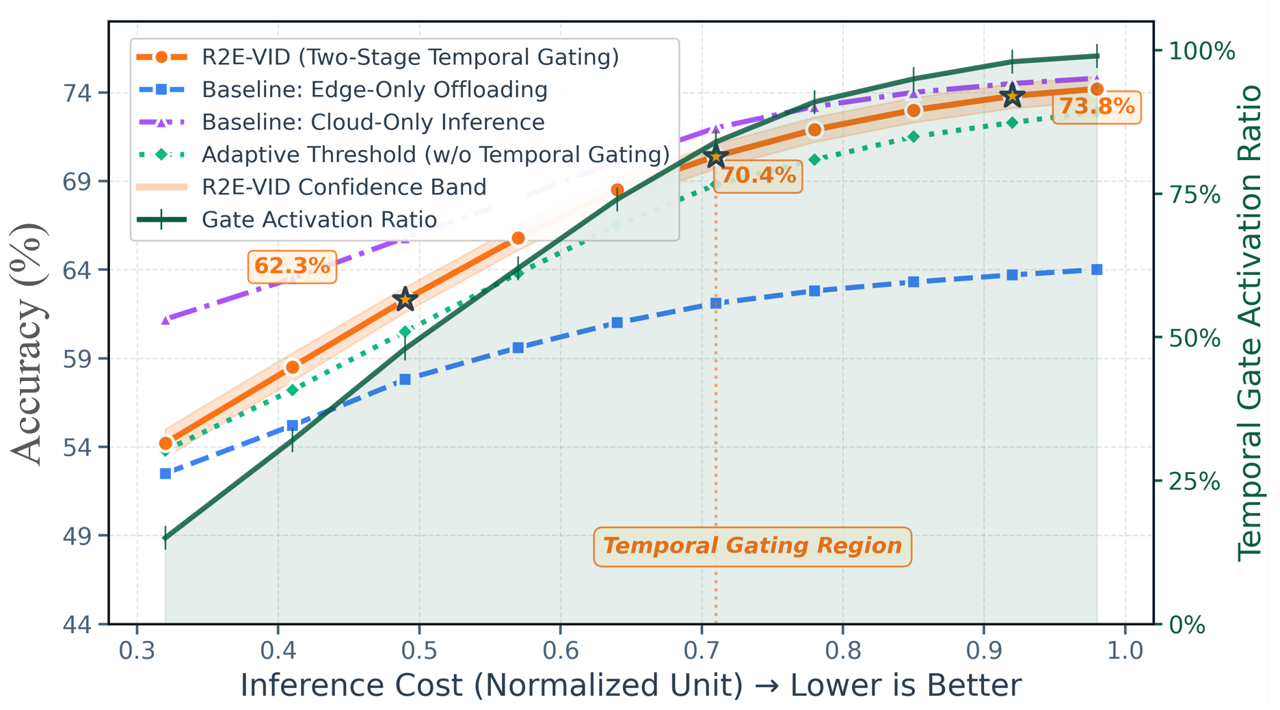}}
        \subfigure[ADE20K]{
    \label{Fig.sub.8.2}
    \includegraphics[width=0.32\textwidth]{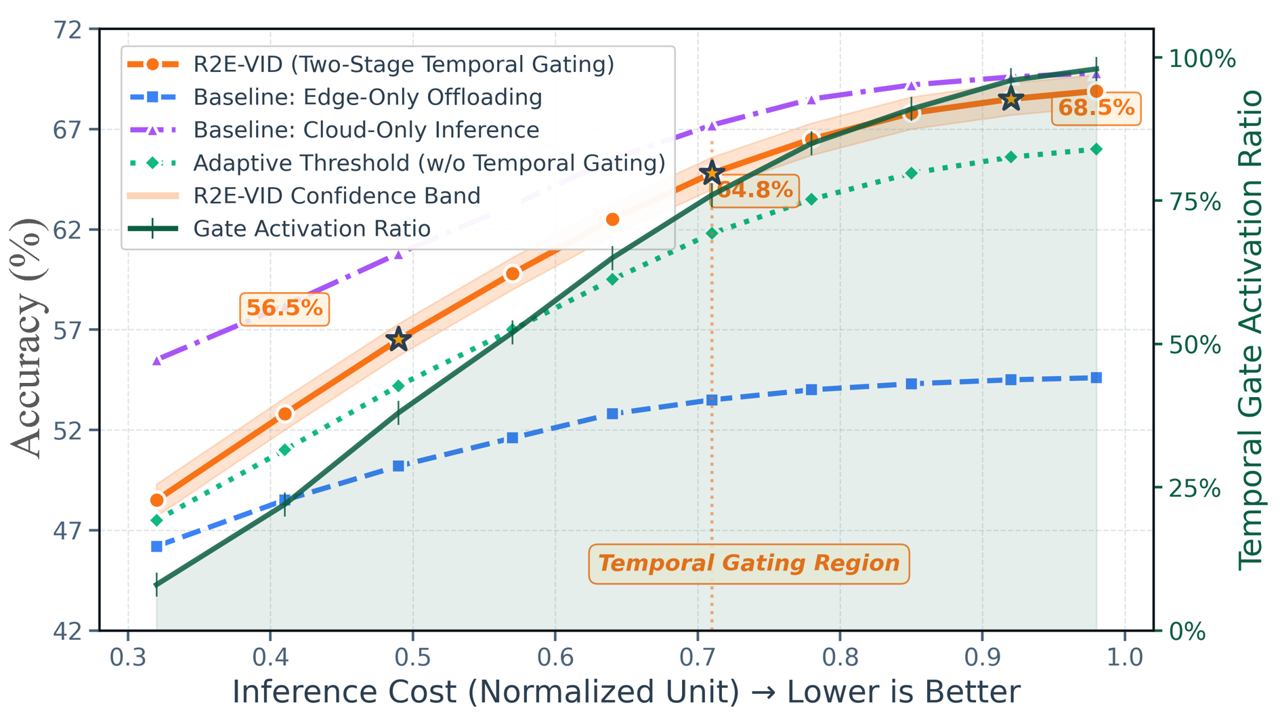}}
    \caption{The accuracy-cost tradeoff analysis under different datasets.}
	\label{figure5}
\end{figure*}

\begin{table*}[th]
\centering
  \caption{The average accuracy comparison results under the object detection task (Take the average of the COCO dataset and UA-DETRAC dataset).}
  \label{tab1}
  \setlength{\tabcolsep}{1.5mm}{
  \begin{tabular}{cccccccccccl}
    \toprule
     \multirow{2}*{Detection Objects} & \multicolumn{5}{c}{Stable Requirements} & \multicolumn{5}{c}{Fluctuating Requirements} & \\
  \cmidrule(lr){2-6}\cmidrule(lr){7-11}
    & $A^2$  & JCAB&RDAP& Sniper& R2E-VID &$A^2$ &  JCAB &RDAP& Sniper& R2E-VID\\
    \midrule
    Cars & 69.54\% & 66.83\% & 63.79\% & 67.26\% & \textbf{69.67\%} & 68.33\% & 65.13\% & 62.11\% & 66.45\% & \textbf{69.54\%}\\
    Buses & 69.73\% & 66.25\% & 62.37\% & 67.11\% & \textbf{69.81\%} &68.14\%  & 65.18\% & 61.16\% & 66.13\% & \textbf{69.16\%}\\
    Motorcycles & 64.06\% & 61.12\% & 57.17\% & 62.21\% &\textbf{64.15\%} & 63.33\%& 60.04\%  & 55.62\% & 61.16\% & \textbf{63.77\%}\\
    Bicycles & 63.15\% & 59.86\% & 55.84\% & 61.03\% &\textbf{63.35\%} & 61.16\% & 59.24\% & 54.23\% & 60.03\% & \textbf{62.86\%}\\
    Persons & 66.31\% & 62.81\% & 58.76\% & 64.12\% &\textbf{66.51\%} & 64.43\%& 62.13\%  & 57.17\% & 63.24\% & \textbf{65.73\%}\\
  \bottomrule
\end{tabular}}
\end{table*}

\subsection{Evaluation Setup}
\subsubsection{Datasets and Implementation Details} To evaluate our solution, we perform object detection experiments on the COCO dataset \cite{lin2014microsoft333333} and the UA-DETRAC dataset \cite{wen2020ua686868}, focusing on detecting objects such as cars, buses, motorcycles, bicycles, and persons. We also conduct semantic segmentation experiments on the ADE20K
dataset \cite{DBLP646464}. The experiments utilize an Intel Xeon Silver 4214R CPU with 64 GB of memory for the cloud server and four NVIDIA Jetson Xavier NX devices with 8 GB of memory each for the edge servers. We employ YOLOv5 series models for object detection tasks and ViT models for semantic segmentation tasks, which include many versions of the model in different sizes. Five models of different sizes are deployed on edge servers and cloud servers, respectively. The size of the model in the cloud is approximately 10 times that of the edge. Baseline methods include (1) \emph{{$A^2$}} \cite{jiang2021joint222}, (2) \emph{JCAB} \cite{wang2020joint111}, (3) \emph{RDAP} \cite{su2022prediction363636}, and  (4) \emph{Sniper} \cite{434343}. We employ the following metrics to assess the performance of various methods. (1) \emph{Accuracy}, (2) \emph{Delay}, (3) \emph{Energy Consumption}, and (4) \emph{Cost}. Where the cost is calculated as a weighted sum of delay and energy consumption.

\subsubsection{Parameter Selection} 
We categorize the accuracy requirements of the task into two types: stable and fluctuating requirements. Stable accuracy requirements are chosen randomly from [0.6, 0.7], and fluctuating accuracy requirements are chosen randomly from [0.5, 0.8]. These ranges represent a wide range of applications \cite{444444}. According to \cite{zhang2018awstream313131}, the bandwidths of the cloud server and edge server are 100 Mbps and 50 Mbps. The powers of the cloud server and edge server are 100 W and 15 W. Finally, five input task resolutions are selected, which are 360p, 540p, 720p, 900p, and 1080p, respectively. The video frame rate range is 10-50 FPS. The number of iterations for the robust optimization algorithm is set to 5000. After preliminary testing, the weighting factor $\beta$ is set to 0.06.

\subsection{Accuracy-Cost Tradeoff Analysis}
To evaluate the accuracy-cost tradeoff enabled by the proposed framework, we vary the cost budget constraint and measure the resulting inference accuracy across three benchmark datasets: COCO, UA-DETRAC, and ADE20K. As shown in Figure~\ref{figure5}(a)–(c), the accuracy consistently increases with the available cost budget across all datasets, demonstrating the framework’s ability to effectively utilize additional resources to enhance inference performance. On COCO, accuracy improves from 59.5\% to 73.8\% as the cost budget increases from 0.5 to 1.0. Similar trends are observed on UA-DETRAC and ADE20K, where accuracy rises from 52.3\% to 60.0\% and from 48.5\% to 56.0\%, respectively.

Notably, the proposed R2E-VID framework consistently outperforms the baseline schemes, including the cloud-only and edge-only configurations, across all datasets and budget settings. By adaptively balancing edge and cloud resources through temporal-aware gating and two-stage robust optimization, R2E-VID achieves a more favorable accuracy-cost tradeoff than either static deployment. In particular, the edge-only baseline suffers from limited computational capacity, resulting in the lowest accuracy across all scenarios, while the cloud-only baseline incurs excessive transmission costs without comparable accuracy gains. In contrast, R2E-VID dynamically allocates video processing tasks between edge and cloud, leveraging the strengths of both to maximize accuracy under given cost constraints. These results validate that the two-stage robust routing mechanism, guided by temporal gating, achieves a flexible and controllable tradeoff between accuracy and processing cost, allowing the system to adapt to varying resource constraints while maintaining stable performance gains over conventional approaches.

\subsection{End-to-End Experimental Results}

\subsubsection{Accuracy Comparison}
The experimental comparison results of accuracy under the object detection task and the semantic segmentation task are presented in Table~\ref{tab1} and Table~\ref{tab2}. R2E-VID achieves an accuracy level comparable to that of $A^2$ under stable requirements and surpasses  $A^2$ when subjected to fluctuating requirements.  R2E-VID has a higher inference accuracy than traditional cloud-only approaches mainly because it can dynamically select the most suitable model version based on the specific context of each task (e.g., resolution, resources, etc.) through a multi-model robust optimization algorithm. Traditional cloud-only methods are limited by insufficient knowledge of scene features and usually select parameters based on global average performance only. In addition, R2E-VID significantly outperforms other edge-cloud methods in terms of accuracy across multiple video tasks, with an average improvement of more than 2\%.

\begin{table}[th]
\centering
\caption{The average accuracy comparison results under semantic segmentation task.}
\label{tab2}
\setlength{\tabcolsep}{0.8mm}{
\begin{tabular}{cccccccccl}
\toprule
\multirow{2}*{Methods} & \multicolumn{2}{c}{Stable Bandwidths} & \multicolumn{2}{c}{Fluctuating Bandwidths} & \\
\cmidrule(lr){2-3}\cmidrule(lr){4-5}
 & MIoU & MPA & MIoU & MPA \\
\midrule
$A^2$  
& $50.95 $ & $79.18 $ 
& $50.69 $ & $79.12 $ \\

JCAB 
& $48.35 $ & $77.14 $ 
& $48.07 $ & $76.82 $ \\

RDAP 
& $45.26 $ & $71.31 $ 
& $44.81 $ & $71.24 $ \\

Sniper 
& $49.13 $ & $77.72 $ 
& $48.82 $ & $77.36$ \\

R2E-VID 
& $\mathbf{51.16}$ & $\mathbf{79.26}$ 
& $\mathbf{51.09}$ & $\mathbf{79.17 }$ \\

\bottomrule
\end{tabular}}
\end{table}

\begin{table}[th]
  \caption{The success rates for meeting the accuracy requirements of task processing.}
  \centering
  \label{tab3}
  \tabcolsep 2pt 
  \begin{tabular}{c|cccccl}
    \toprule
    \multicolumn{2}{c}{Methods}&$A^2$&JCAB &RDAP& Sniper &Ours\\
    \midrule
    \multirow{3}*{COCO} 
    &Stable &  94\%  & 87\% & 86\% & 88\% & \textbf{97\%} \\
    &Fluctuating  &  90\% & 83\% & 81\% & 85\% & \textbf{96\%}\\
       \midrule
    \multirow{3}*{UA-DETRAC } 
    &Stable &91\%  & 84\% & 84\% & 86\% & \textbf{95\%} \\
    &Fluctuating &  88\% & 81\% & 79\% & 82\% & \textbf{94\%}\\
           \midrule
    \multirow{3}*{ADE20K} 
    &Stable &89\%  & 80\% & 79\% & 81\% & \textbf{92\%} \\
    &Fluctuating &  88\% & 75\% & 74\% & 78\% & \textbf{91\%}\\
  \bottomrule
\end{tabular}
\end{table}

    
    
    
    

Next, we gauge its success rate in meeting the predetermined task accuracy requirements. Specifically, if the final inference accuracy of a given task aligns with the specified input requirements, it is deemed successful; otherwise, it is considered a failure. As shown in Table~\ref{tab3}, we observe that R2E-VID achieves the maximum success rate across different accuracy requirements. And the advantage is more pronounced under fluctuating requirements. Our method still achieves an average success rate of over 91\% under dynamic conditions, which is 6\% to 17\% higher than other methods. This result shows that it can respond to diverse accuracy requirements by adjusting the task resolution and model version. This underscores the adaptive capabilities of R2E-VID in handling variable accuracy requirements.


\begin{figure}[th]
	\centering 
    \subfigure[Delay]{
    \label{Fig.sub.7.1}
    \includegraphics[width=0.23\textwidth]{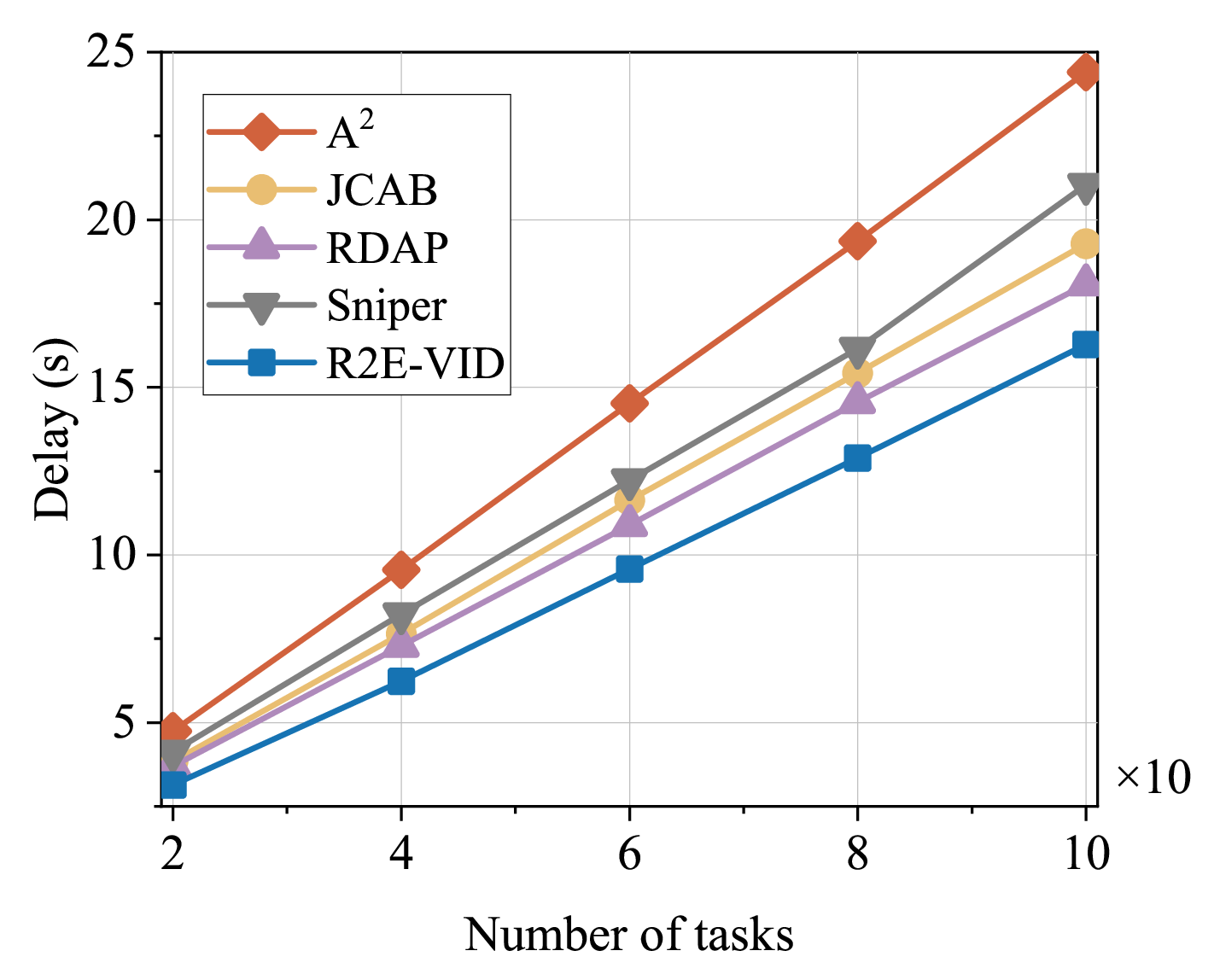}}
    \subfigure[Energy Consumption]{
    \label{Fig.sub.7.2}
    \includegraphics[width=0.23\textwidth]{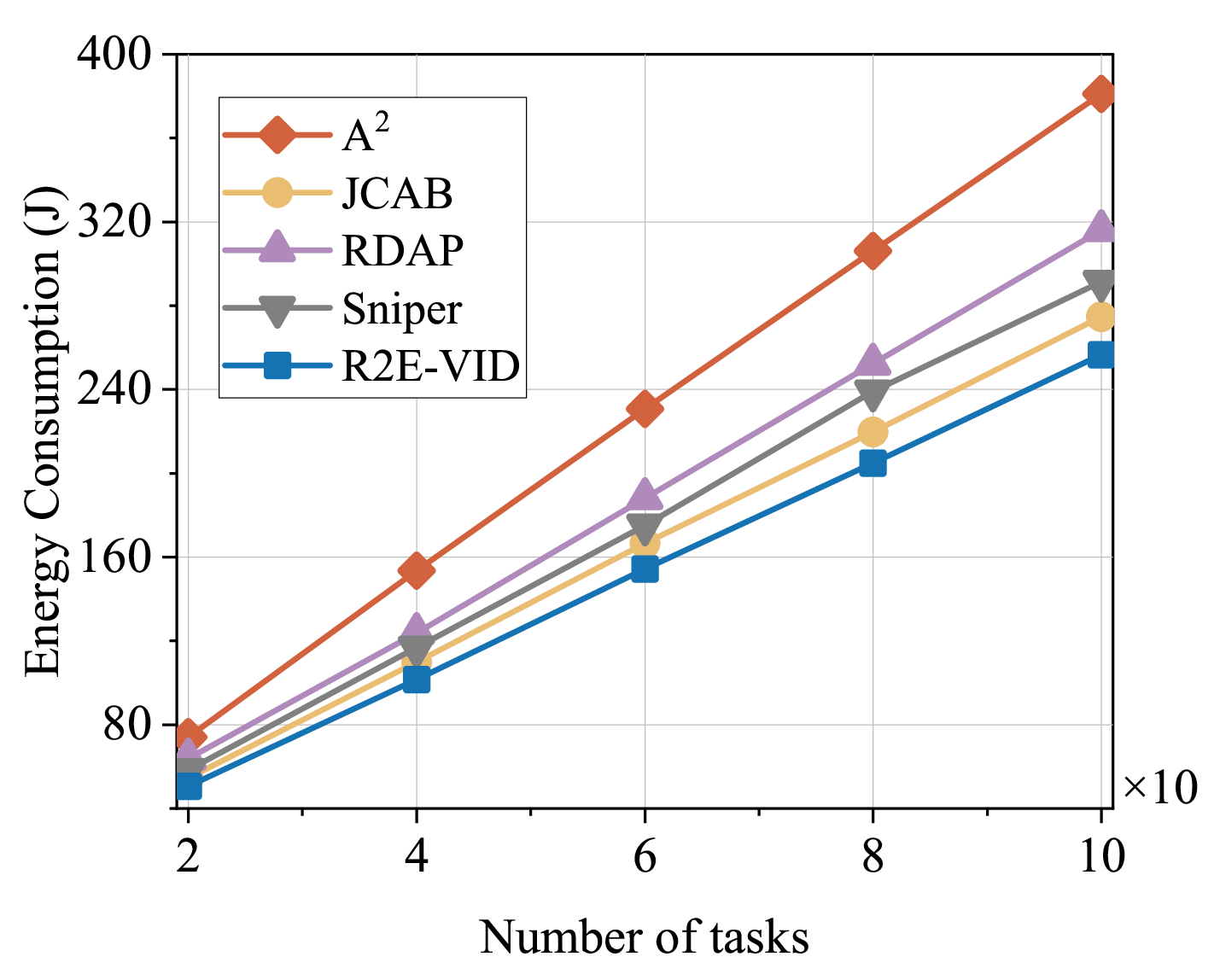}}
    \caption{The comparative results of different methods under the COCO dataset.}
	\label{figure6}
\end{figure}

\begin{figure}[th]
	\centering 
    \subfigure[Delay ]{
    \label{Fig.sub.7.1}
    \includegraphics[width=0.23\textwidth]{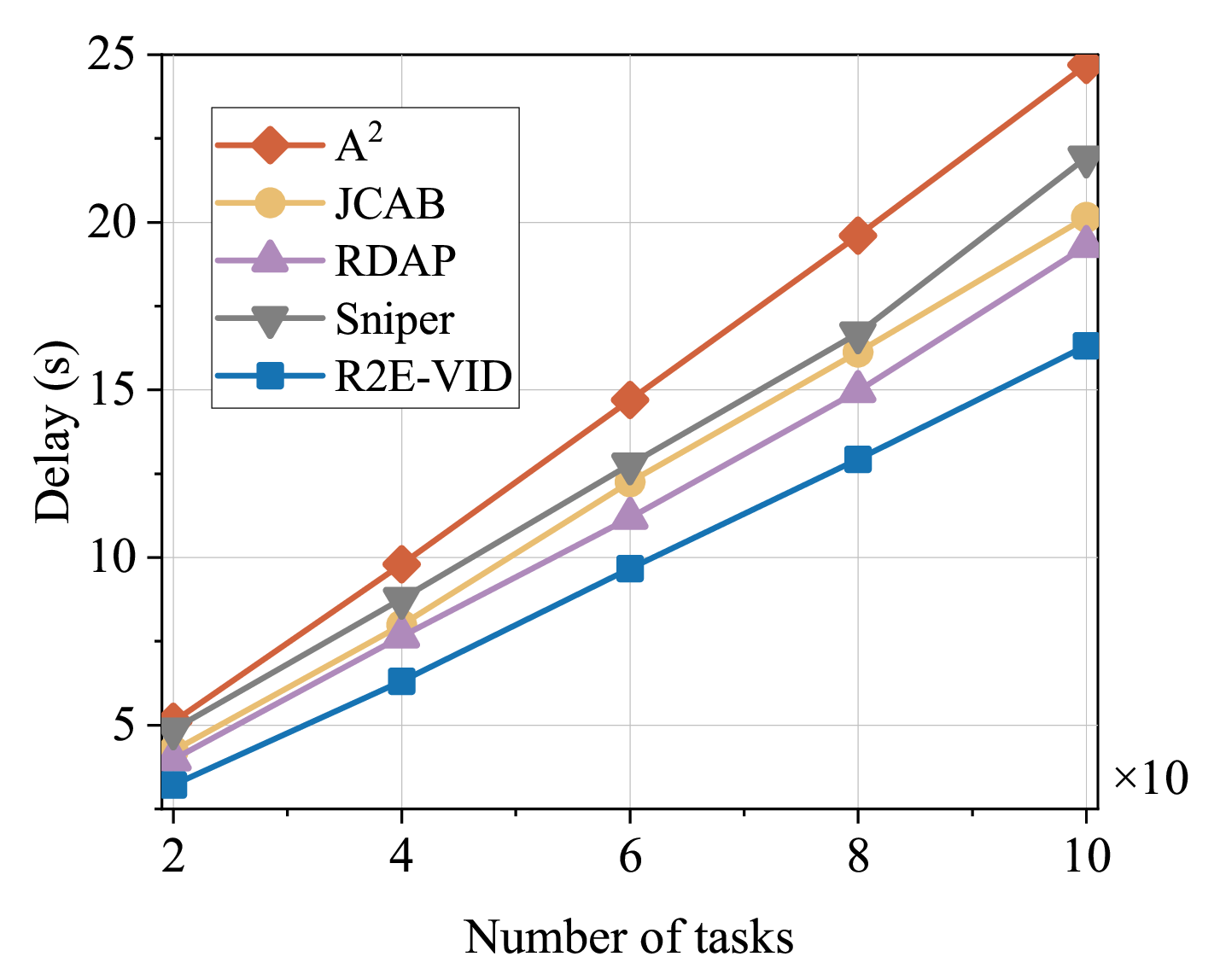}}
    \subfigure[Energy Consumption]{
    \label{Fig.sub.7.2}
    \includegraphics[width=0.23\textwidth]{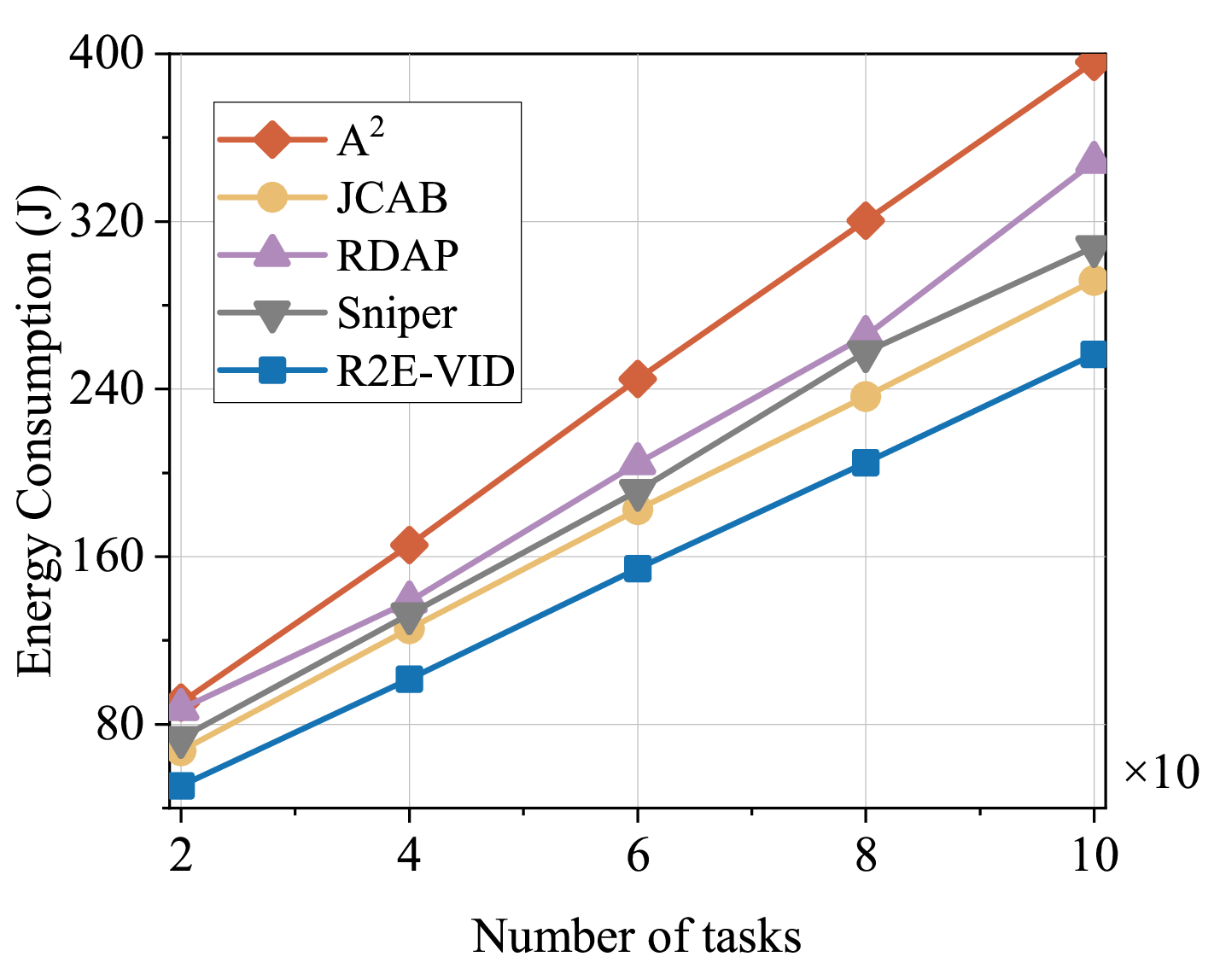}}
    \caption{The comparative results of different methods under the UA-DETRAC dataset.}
	\label{figure7}
\end{figure}

\begin{figure}[th]
	\centering 
    \subfigure[Delay]{
    \label{Fig.sub.7.1}
    \includegraphics[width=0.23\textwidth]{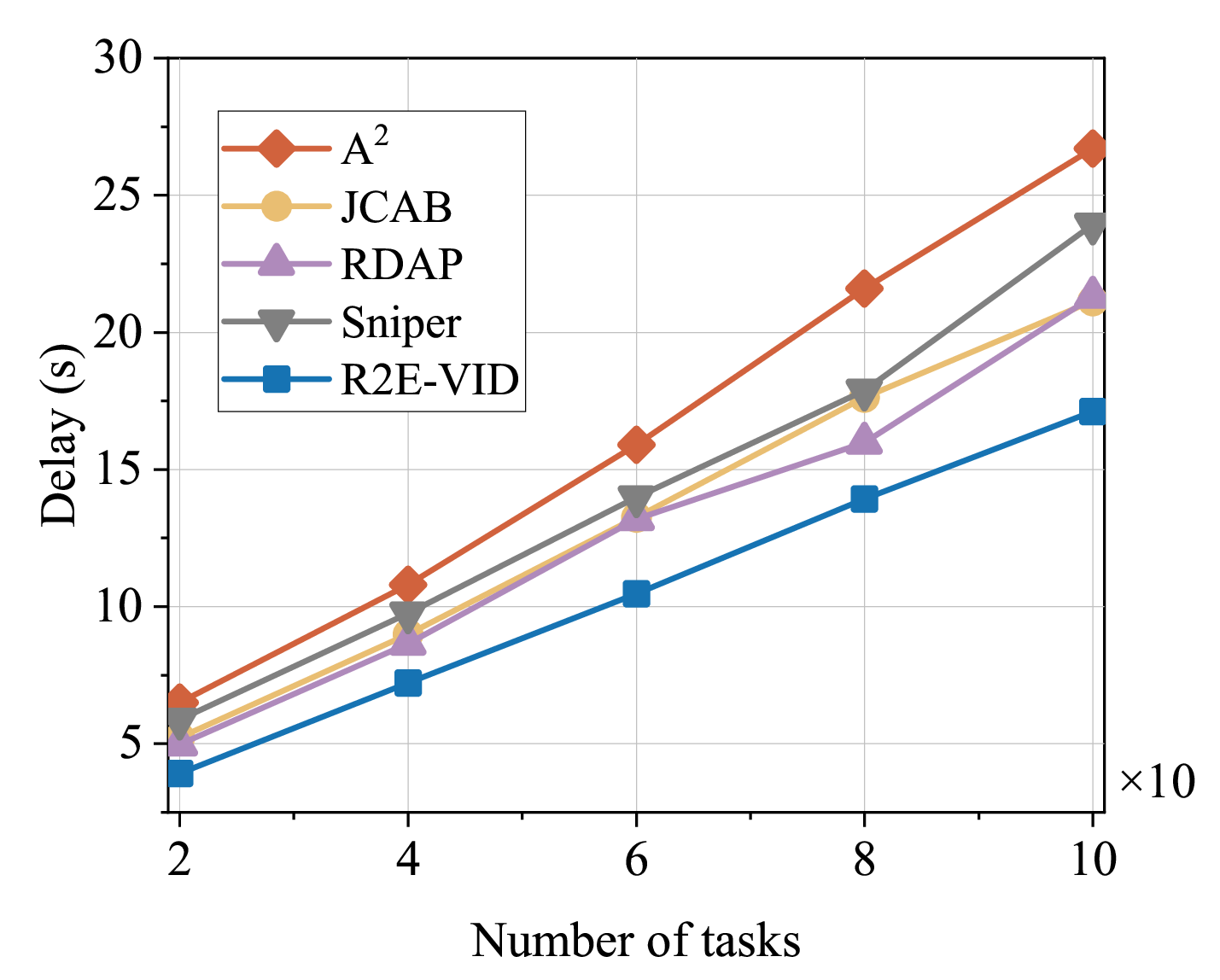}}
    \subfigure[Energy Consumption]{
    \label{Fig.sub.7.2}
    \includegraphics[width=0.23\textwidth]{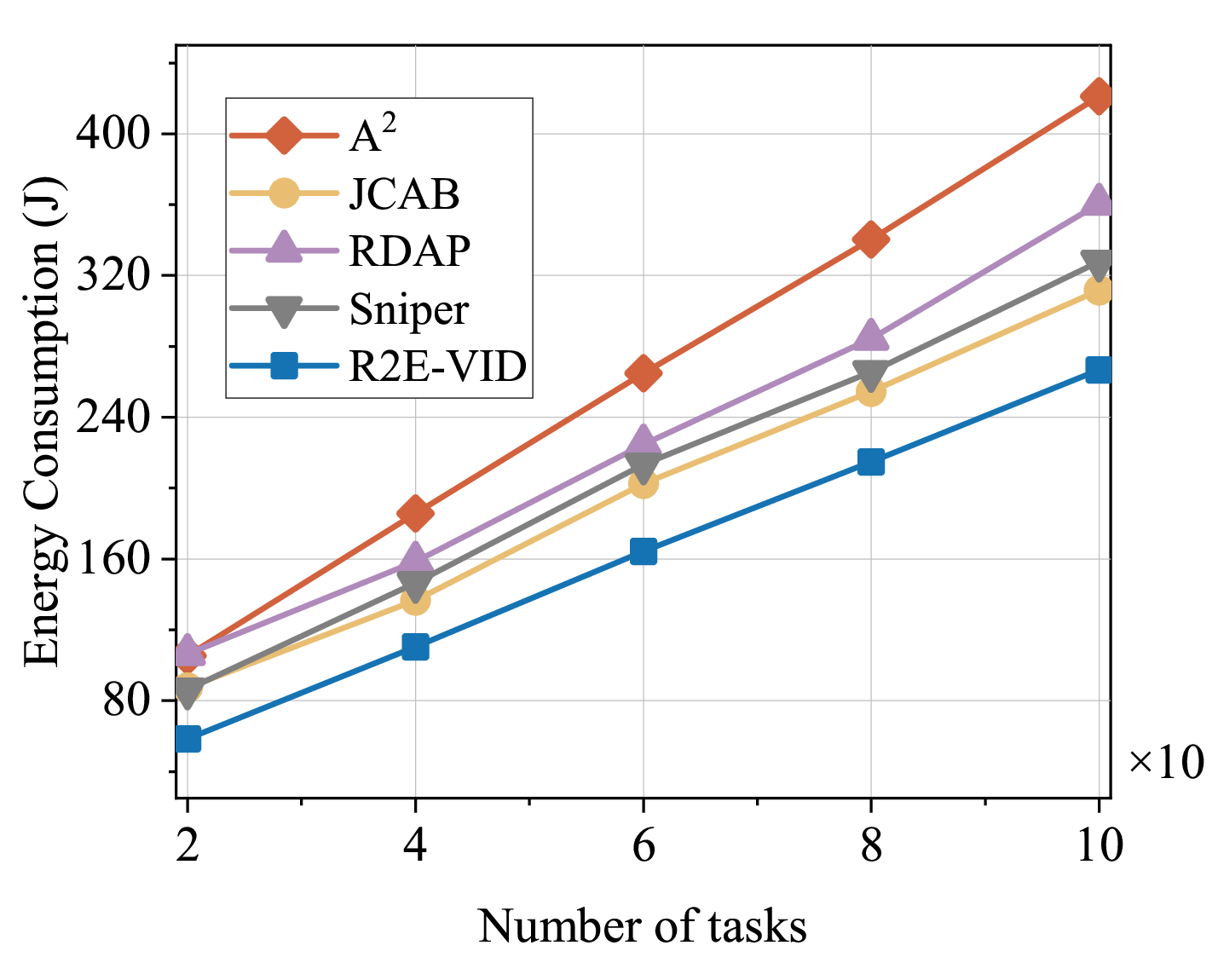}}
    \caption{The comparative results of different methods under the ADE20K dataset.}
	\label{figure8}
\end{figure}

\subsubsection{Robustness to Changes in the Number of Tasks}
To assess overall performance, we evaluate all methods under varying task volumes, with results averaged over stable and fluctuating accuracy requirements. As shown in Figure~\ref{figure6},  Figure~\ref{figure7}, and Figure~\ref{figure8}, R2E-VID consistently achieves the lowest delay, and its advantage widens as the number of tasks increases. This improvement stems from two key mechanisms: (1) adaptive resolution and frame-rate selection, which reduce transmission time and balance edge–cloud workloads, and (2) robust model versioning, which preserves accuracy while minimizing unnecessary computation. In terms of energy consumption, R2E-VID again delivers the best performance across all settings, showing both lower delay and reduced energy usage under increasing task loads. These results demonstrate that R2E-VID effectively balances accuracy, delay, and energy through joint edge–cloud optimization and adaptive video configuration. The gains become more pronounced with larger workloads, highlighting the scalability and robustness of the framework in dynamic, real-world environments.

\begin{figure}[th]
	\centering 
    \subfigure[COCO]{
    \label{Fig.sub.9.1}
    \includegraphics[width=0.45\textwidth]{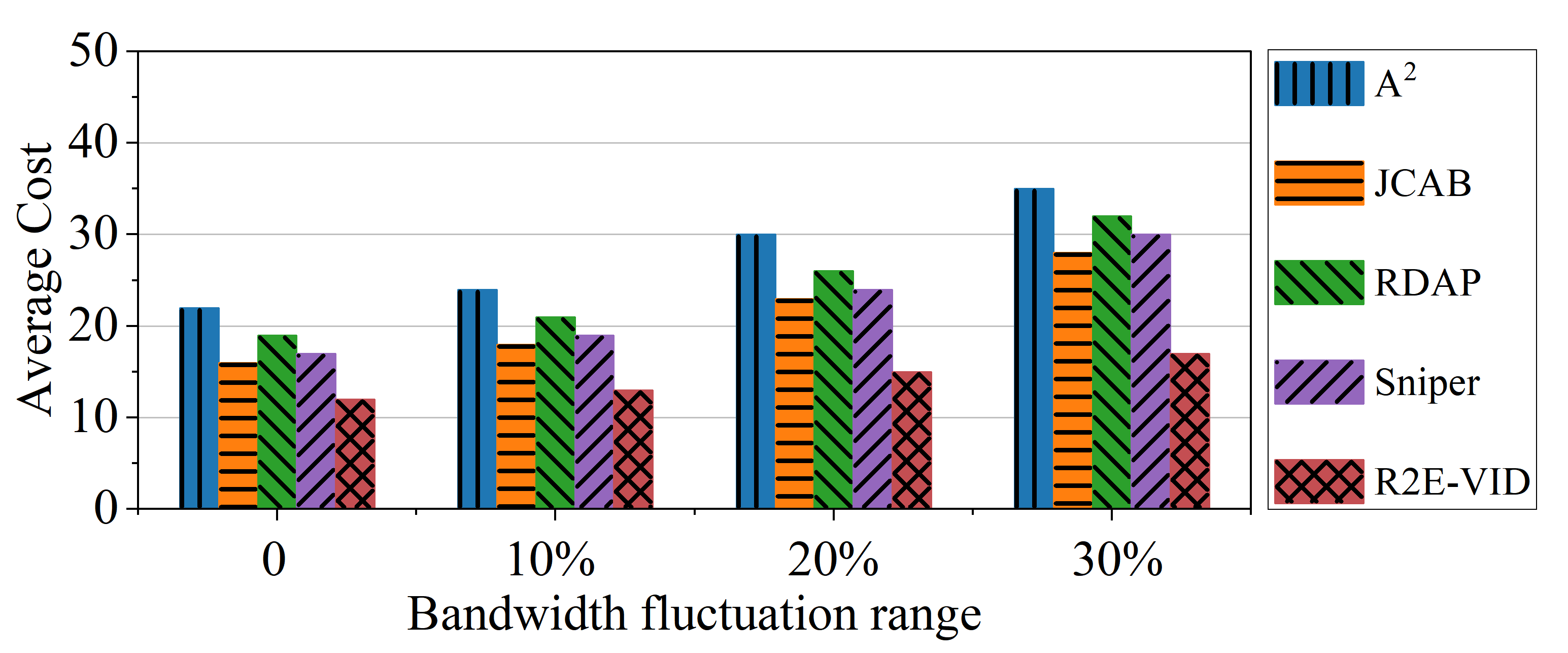}}
    
    \subfigure[UA-DETRAC]{
    \label{Fig.sub.9.2}
    \includegraphics[width=0.45\textwidth]{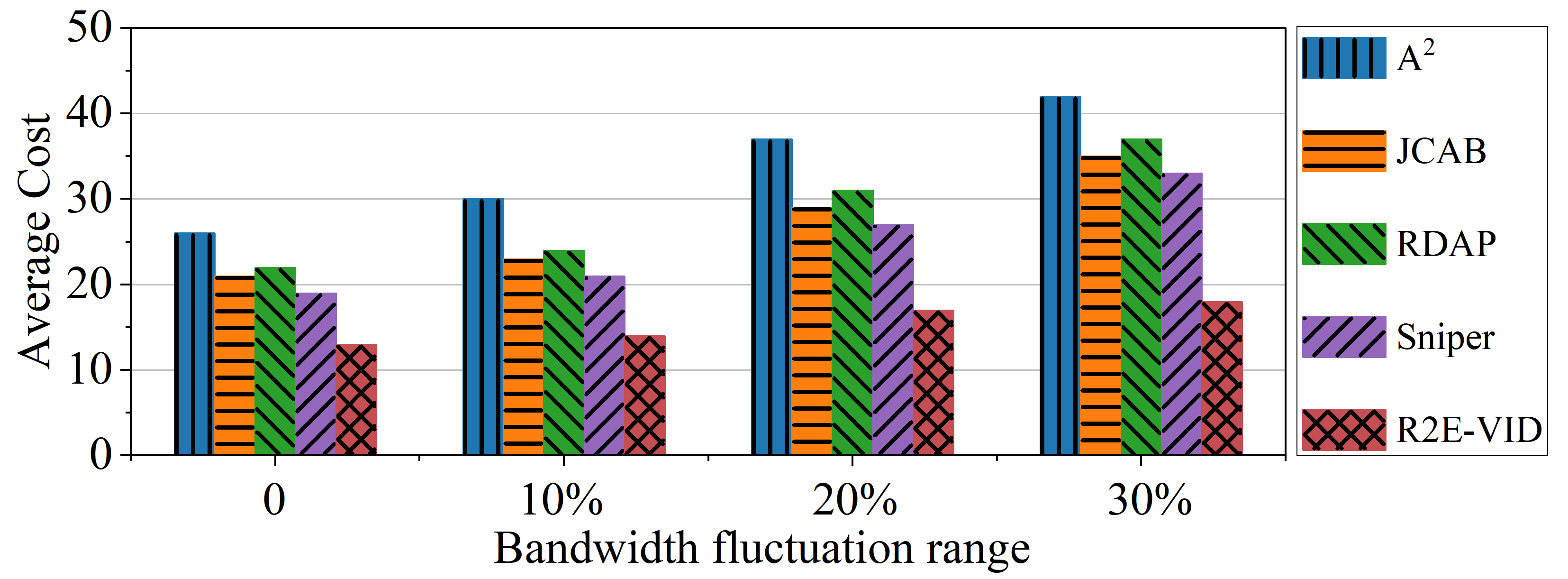}}

        \subfigure[ADE20K]{
    \label{Fig.sub.9.3}
    \includegraphics[width=0.45\textwidth]{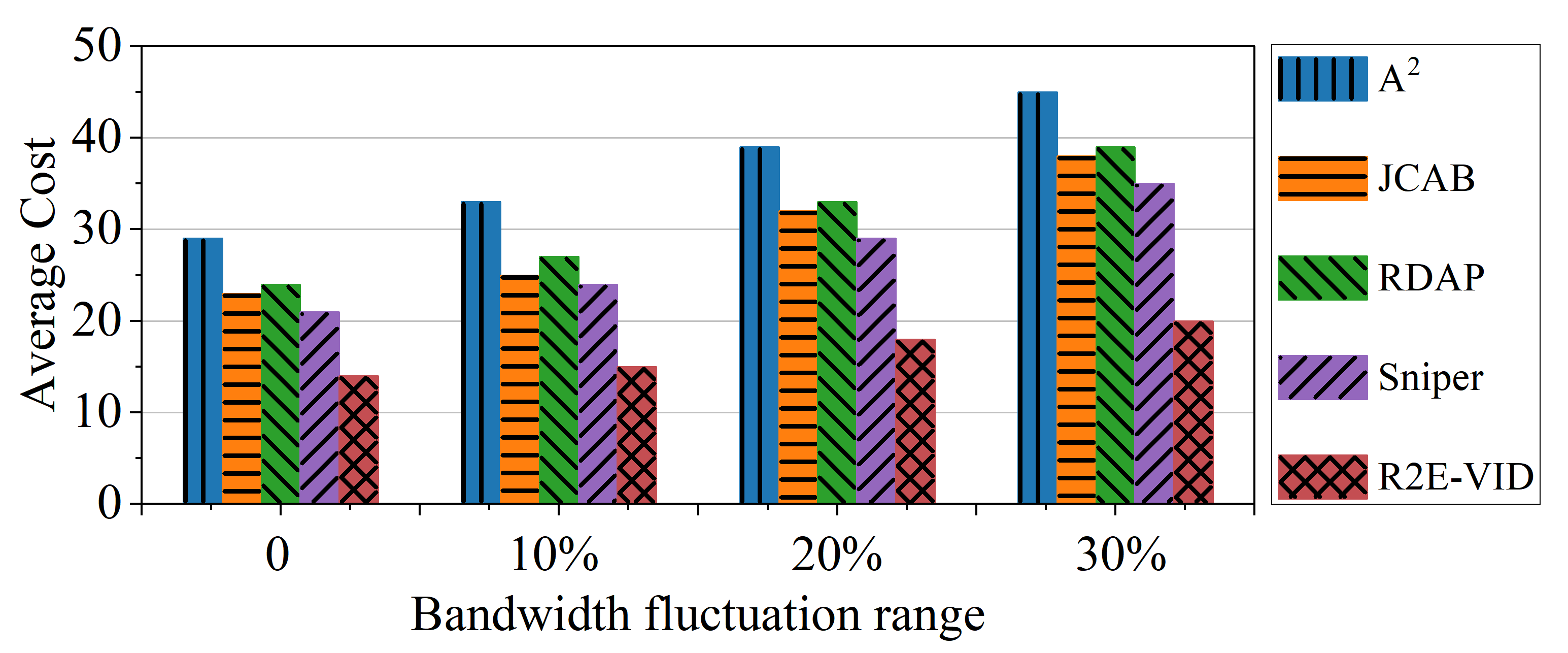}}
    \caption{The average cost comparison of different methods under dynamic bandwidths.}
	\label{figure9}
\end{figure}

\subsubsection{Robustness to Dynamic Network}
To further assess robustness under real-world conditions, we evaluate all methods under dynamic network environments, where the bandwidth fluctuates within {0, 10\%, 20\%, 30\%}. All other settings remain unchanged. The results in Figure~\ref{figure8} show the total cost on COCO (Figure~\ref{Fig.sub.9.1}), UA-DETRAC (Figure~\ref{Fig.sub.9.2}), and ADE20K (Figure~\ref{Fig.sub.9.3}). Across all fluctuation ranges, R2E-VID consistently achieves the lowest cost. As bandwidth variation increases, the costs of competing methods rise sharply, whereas R2E-VID degrades much more slowly. This demonstrates the strong resilience enabled by its adaptive two-stage robust optimization design. Quantitatively, R2E-VID reduces cost by 35\%–45\% compared with JCAB, RDAP, and Sniper, and by over 60\% relative to the cloud-only solution $A^2$. These gains highlight the framework’s effectiveness in managing uncertainty and resource variability. Overall, R2E-VID achieves a well-balanced tradeoff between accuracy and cost, maintaining superior end-to-end performance even under highly dynamic network conditions, and confirming its applicability to diverse real-world edge–cloud video inference scenarios.

\subsection{Ablation Studies}
\begin{figure}[t!]
	\centering 
		\includegraphics[width=1\linewidth]{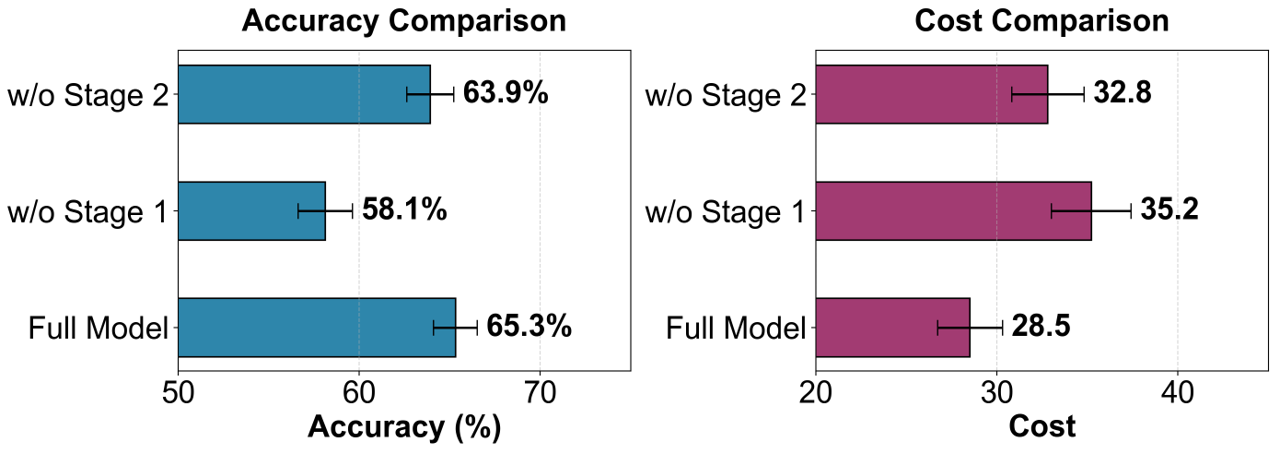}
	\caption{The ablation studies of two-stage robust optimization.}
	\label{figure10}
\end{figure}

To further validate the effectiveness and necessity of each component in the proposed framework, we conduct ablation experiments on the two-stage robust optimization strategy. The evaluation focuses on two key performance metrics: average inference accuracy and average processing cost. We systematically disable individual components, specifically \textbf{Stage 1} and \textbf{Stage 2}, to observe their respective impacts. As shown in Figure~\ref{figure10}, removing \textbf{Stage 1} leads to a noticeable drop in accuracy, from 65.3\% (full model) to 58.1\%, corresponding to a relative reduction of approximately 11\%. This highlights the critical role of \textbf{Stage 1} in enabling temporal-aware video configuration and edge–cloud partitioning, which together facilitate adaptive resolution selection and routing decisions. In contrast, disabling \textbf{Stage 2} results in a more moderate accuracy decline to 63.9\%, suggesting that \textbf{Stage 2} primarily contributes to robustness rather than directly determining the accuracy baseline. In terms of processing cost, the full model achieves the lowest cost at 28.5\%. Removing either stage leads to a substantial cost increase: 32.8\% when \textbf{Stage 2} is ablated and 35.2\% when Stage 1 is ablated. These increases of approximately 15\% and 23\% demonstrate that both stages play a complementary role in maintaining resource efficiency. The results confirm that the joint operation of both stages is essential for achieving an optimal balance between inference accuracy and processing cost.


\section{Conclusion} \label{Conclusion}
In this paper, we presented R2E-VID, a two-stage robust routing framework designed to enable cost-efficient and accuracy-aware video inference in heterogeneous edge–cloud systems. Addressing the inherent complexity of jointly optimizing video configuration, routing decisions, and multi-model selection, our framework decouples the problem into two tractable yet tightly coordinated stages. The first stage employs a temporal gating–based adaptive configuration module that captures motion dynamics and temporal consistency in the video stream, allowing the system to adjust resolution, frame rate, and edge–cloud partitioning in real time. The second stage further refines the decision space through robust multi-model selection, ensuring reliable performance under fluctuating network conditions and uncertain resource availability. Experimental results demonstrate that our method can effectively achieve the tradeoff between accuracy and cost. Compared to other baseline methods, it can reduce the cost by 35\%-60\% on public datasets and ensure accuracy.

\bibliographystyle{ACM-Reference-Format}
\bibliography{sample-base}

\appendix

\end{document}